\documentclass{aa}
\usepackage{graphicx}
\usepackage{txfonts}
\newcommand{\degree}{^\circ}

\begin{document} 
\title{
Searching for Galactic hidden gas through interstellar scintillation:
Results from a test with the NTT-SOFI detector
\thanks{This work is based on observations made at the
European Southern Observatory, La Silla, Chile.}
}

\author{
F.~Habibi\inst{1}$^,$\inst{2},
M.~Moniez\inst{1},
R.~Ansari\inst{1},
S.~Rahvar\inst{2}
}
\institute{
Laboratoire de l'Acc\'{e}l\'{e}rateur Lin\'{e}aire,
{\sc IN2P3-CNRS}, Universit\'e de Paris-Sud, B.P. 34, 91898 Orsay Cedex, France
\and
Department of Physics, Sharif University of Technology
PO Box 11365-9161, Tehran, Iran
}

\offprints{M. Moniez, \email{ moniez@lal.in2p3.fr}}

\date{Received 22/07/2009, accepted 17/10/2010}
%

\abstract
{}{
Stars twinkle because their light propagates through the atmosphere.
The same phenomenon is expected at a longer time scale
when the light of remote stars
crosses an interstellar molecular cloud, but it has never been
observed at optical wavelength.
In a favorable case, the light of a background star can be subject to
stochastic fluctuations on the order of a few percent at a characteristic
time scale of a few minutes.
Our ultimate aim is to discover or exclude these scintillation effects
to estimate the contribution of molecular hydrogen
to the Galactic baryonic hidden mass.
This feasibility study is a pathfinder toward an observational
strategy to search for scintillation, probing the sensitivity of
future surveys and estimating the background level.
}
{
We searched for scintillation induced by molecular gas in visible
dark nebulae as well as by hypothetical halo clumpuscules of
cool molecular hydrogen ($\mathrm{H_2-He}$) during two nights.
We took long series of 10s infrared exposures with the ESO-NTT
telescope
toward stellar populations located behind visible nebulae
and toward the Small Magellanic Cloud (SMC).
We therefore searched for stars exhibiting stochastic flux variations
similar to what is expected from the scintillation effect.
According to our simulations of the scintillation process,
this search should allow one to detect (stochastic)
transverse gradients of column density
in cool Galactic molecular clouds of order of
$\sim 3\times 10^{-5}\,\mathrm{g/cm^2/10\,000\,km}$.
}
{
We found one light-curve that is compatible with a strong
scintillation effect through a turbulent structure characterized by
a diffusion radius $R_{diff}<100\, km$ in the B68 nebula.
Complementary observations are needed to
clarify the status of this candidate, and no firm conclusion
can be established from this single observation.
We can also infer limits on the existence of
turbulent dense cores (of number density $n>10^9\, cm^{-3}$)
within the dark nebulae.
Because no candidate is found toward the SMC,
we are also able to establish
upper limits on the contribution
of gas clumpuscules to the Galactic halo mass.
}
{
The limits
set by this test do not seriously constrain the known models, but we show that
the short time-scale monitoring for a few $10^6 star\times hour$
in the visible band
with a $>4$ meter telescope and a fast readout camera should
allow one to quantify the contribution of turbulent molecular gas to
the Galactic halo.
The LSST (Large Synoptic Survey Telescope) is perfectly suited
for this search.
}


\titlerunning{Interstellar Optical Scintillation}
\authorrunning{Habibi, Moniez, Ansari, Rahvar}
\maketitle

\section{Introduction}
The present study was made to explore the feasibility
of the detection of scintillation effects
through nebulae and through hypothetical cool molecular hydrogen
($\mathrm{H_2-He}$) clouds.
Considering the results of baryonic compact massive objects
searches through microlensing (\cite{ErosLMCfinal}; \cite{OgleSMC}; 
\cite{macho2000LMC}; see also the review of \cite{reviewMoniez}),
these clouds should now be seriously considered as a possible major component
of the Galactic hidden matter.
It has been suggested that a hierarchical structure of cold $\mathrm{H_2}$
could fill the Galactic thick disk (\cite{fractal1} 1994; \cite{fractal2})
or halo (\cite{Jetzer1}; \cite{Jetzer2}), 
providing a solution for the Galactic hidden matter problem.
This gas should form transparent ``clumpuscules'' of
$\sim 30\,\mathrm{AU}$ size,
with an average density of $10^{9-10} \mathrm{cm^{-3}}$, an average
column density of $10^{24-25}\,\mathrm{cm^{-2}}$, and a
surface filling factor of $\sim 1\%$.
The detection of these structures thanks to the scintillation of
background stars would have a major impact on the Galactic dark matter
question.

The OSER project (Optical Scintillation by Extraterrestrial Refractors)
is proposed to search for scintillation of extra-galactic sources
through these Galactic -- disk or halo -- transparent $\mathrm{H_2}$ clouds.
This project should allow one to detect (stochastic)
transverse gradients of column density
in cool Galactic molecular clouds on the order of
$\sim 3\times 10^{-5}\,\mathrm{g/cm^2/10\,000\,km}$.
The technique can also be used for the nebulae science.
The discovery of scintillation through visible nebulae should
indeed open a new window to investigate their structure.

The feasibility study described here concerns the search for scintillation
through known dark nebulae such as B68 (also identified as LDN57), cb131
(also identified as B93 and L328), through a nebula within the
Circinus complex (hereafter called Circinus nebula), and also includes
a test for hidden matter search toward the SMC.
As discussed below, we did not think it very likely to discover
a signal, and the main purpose of the test was to predict the sensitivity
of a future optical survey from the measurement of the signal
sensitivity in infrared and from the estimate of the variable
star background level.

\section{The scintillation process}
\subsection{Basics}
\begin{figure*}[!ht]
\begin{center}
\includegraphics[width=12.cm]{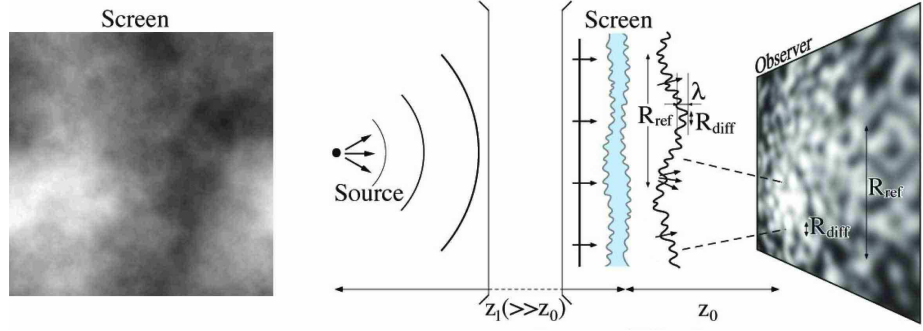}
\end{center}
\caption[] 
{\it
Left: a 2D stochastic phase screen (gray scale) from
a simulation of gas affected by Kolmogorov-type turbulence.

Right: the illumination pattern from a point source (left) after crossing
such a phase screen.
The distorted wavefront produces structures at scales of
$\sim R_{diff}(\lambda)$
and $R_{ref}(\lambda)$ on the observer's plane.
}
\label{front}
\end{figure*}

Refraction through
an inhomogeneous transparent cloud (hereafter called screen),
which is described by a 2D phase delay function $\phi(x,y)$ in the
plane transverse to the line of sight,
distorts the wave-front of incident electromagnetic waves
(Fig. \ref{front})(\cite{Moniez}).
The luminous amplitude in the observer's plane after propagation
is described by the Huygens-Fresnel diffraction theory.
For a {\it point-like, monochromatic} source,
the intensity in the observer's plane is affected by
interferences which take on the speckle aspect
in the case of stochastic inhomogeneities.
At least two distance scales characterize this speckle,
which are related to the wavelength $\lambda$, to the
distance of the screen $z_0$,
and to the statistical characteristics of the stochastic
phase delay function $\phi$:
\begin{itemize}
\item
{\bf The diffusion radius} $R_{diff}(\lambda)$ of the screen, defined as the
separation in the screen transverse plane
for which the root mean square of the phase delay difference
at wavelength $\lambda$ is 1 radian~(\cite{narayan92}).
Formally, $R_{diff}(\lambda)$ is given in a way that
$<(\phi(x'+x,y'+y)-\phi(x',y'))^2>=1$ where $(x',y')$
spans the entire screen plane and $(x,y)$ satisfies
${\sqrt{x^2+y^2}=R_{diff}(\lambda)}$.
The diffusion radius characterizes the structuration of the inhomogeneities
of the cloud, which are related to the turbulence.
As demonstrated in Appendix A, assuming that the cloud turbulence
is isotropic and is described
by the Kolmogorov theory up to the largest scale (the cloud's width $L_z$),
$R_{diff}$ can be expressed as
\begin{equation}
R_{diff}(\lambda)=263\, km\times
\left[\frac{\lambda}{1\mu m}\right]^{\frac{6}{5}}
\left[\frac{L_z}{10\ AU}\right]^{-\frac{1}{5}}
\left[\frac{\sigma_{3n}}{10^{9}\, cm^{-3}}\right]^{-\frac{6}{5}}\!\! ,
\label{expression-rdiff}
\end{equation}
where $\sigma_{3n}$ is the molecular number density dispersion
within the cloud. In this expression, we assume that the average
polarizability of the molecules in the medium is
$\alpha = 0.720\times 10^{-24} cm^3$, corresponding to a mixing of
$76\%$ of ${\rm H_2}$ and $24\%$ of He by mass.
\item
{\bf The refraction radius}
\begin{equation}
\label{Rref}
R_{ref}(\lambda)=\frac{\lambda z_0}{R_{diff}} \sim
30,860\, km\left[\frac{\lambda}{1\mu m}\right]\left[\frac{z_0}{1\, kpc}\right]\left[\frac{R_{diff}(\lambda)}{1000\, km}\right]^{-1},
\end{equation}
is the size in the observer's plane
of the diffraction spot from a patch
of $R_{diff}(\lambda)$ in the screen's plane.
\item
In addition, long scale structures of the screen can possibly induce
local focusing/defocusing configurations that produce long time-scale
intensity variations.
\end{itemize}

\subsection{Expectations from simulation: intensity modulation, time
scale}
\begin{figure}[h]
\centering
\parbox{8cm}{
\includegraphics[width=8cm]{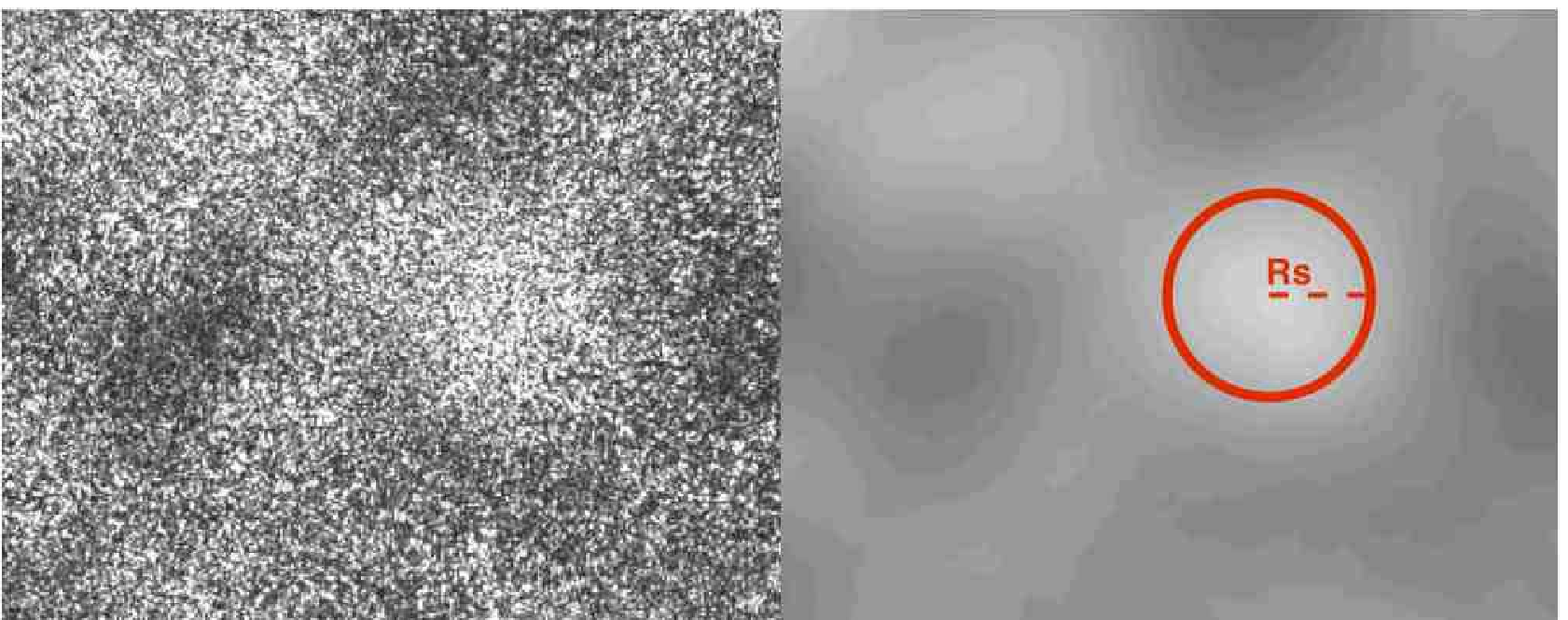}
\includegraphics[width=8cm]{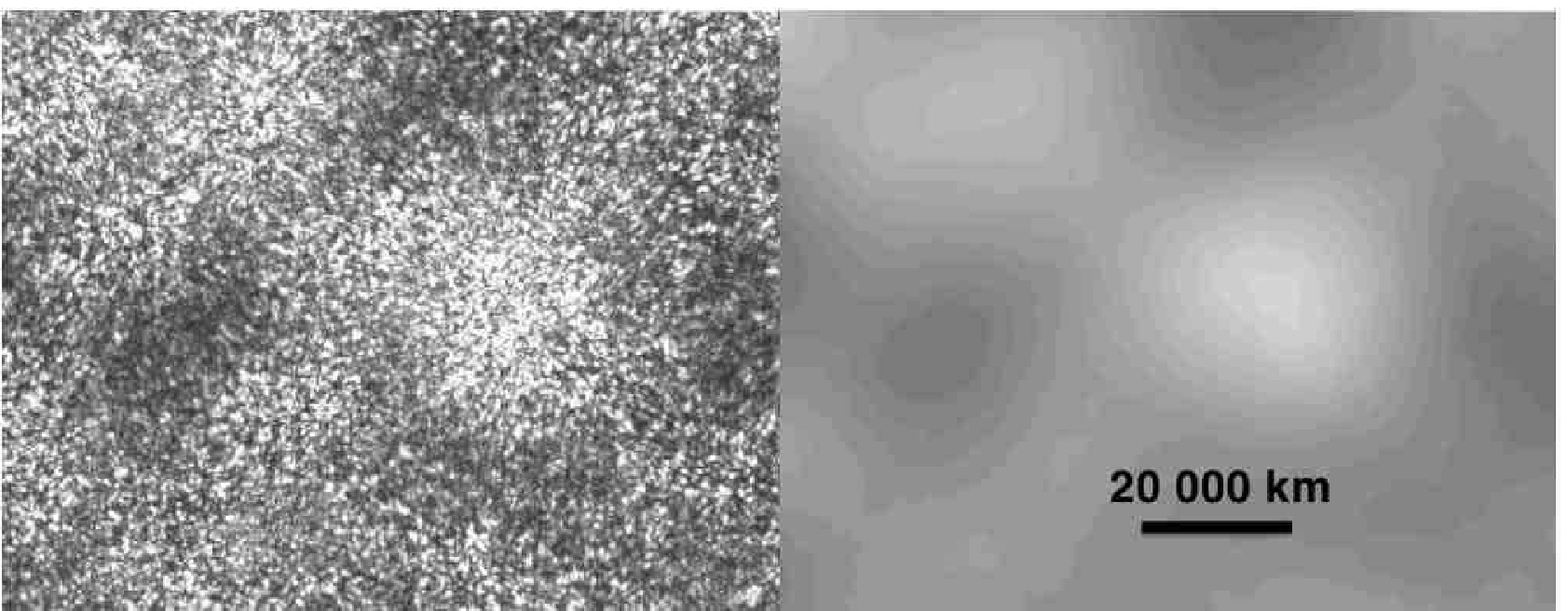}
}
\caption[]{\it
Simulated
illumination map at $\lambda=2.16\mu m$ on Earth from a point source (up-left)-
and from a K0V star ($r_s=0.85R_{\odot}$, $M_V=5.9$) at $z_1=8\, kpc$ (right).
The refracting cloud is assumed to be at $z_0=160\, pc$
with a turbulence parameter $R_{diff}(2.16\mu m)=150\, km$.
The circle shows the projection of the stellar disk (with radius
$R_S=r_s\times z_0/z_1$).
The bottom maps are illuminations in the $K_s$ wide band
($\lambda_{central}=2.162\mu m$, $\Delta\lambda = 0.275\mu m$).
\label{ecran}}
\end{figure}

After crossing an inhomogeneous cloud described by the Kolmogorov turbulence
law (Fig.\ref{front}, left), the light from a {\it monochromatic point}
source produces an illumination
pattern on Earth made of speckles of a size $R_{diff}(\lambda)$ within
larger structures of a size $R_{ref}(\lambda)$ (see Fig.\ref{ecran}, up-left).
The illumination pattern from a real stellar source of radius $r_s$ is
the convolution of the point-like intensity pattern with the projected
intensity profile of the source (projected radius
$R_S=r_s\times z_0/z_1$) (Fig.\ref{ecran}, up-right).
$R_S$ is then another characteristic spatial scale that affects the
illumination pattern from a stellar (not point-like) source.

We simulated these illumination patterns that are caused by the diffusion
of stellar source light through various turbulent media as follows:
\begin{itemize}
\item
We first simulated 2D turbulent screens as stochastic
phase delay functions $\Phi(x,y)$, according to the Kolmogorov law
\footnote{For the Kolmogorov turbulence, the 3D spectral
density is a power law relation with exponent $\beta = 11/3$
(see appendix A). We also explored power laws with different $\beta$
values and found the same general features.};
series of screens were generated at $z_0=125\, pc$ with
$100\, km<R_{diff}<350\, km$.
\item
We then computed the expected illumination patterns
at $\lambda=2.162\mu m$ (central wavelength for $K_s$
filter) from diffused background {\it point-like monochromatic} sources
located at $z_1=1\, kpc$ using the Fast Fourier Transform (FFT)
technique (see Fig.\ref{ecran} up-left).
Illumination patterns for other wavelengths and
geometrical configurations were deduced by simple scaling.
In particular,
configurations compatible with the structure of the $H_2$
clumpuscules of (\cite{fractal1} 1994) were produced for the $J$ passband
($R_{diff}(1.25\mu m)\gtrsim 17\, km$, corresponding to clumpuscules
with  $L_z=30AU$ and $\sigma_{3n}<n_{max}=10^{10} cm^{-3}$).
\item
We derived series of patterns from diffused {\it extended} stellar sources
(radii $0.25 R_{\odot}<r_s<1.5 R_{\odot}$) by convolution
(Fig.\ref{ecran} top-right).
\item
Finally, we co-added the patterns obtained for the central and
the two extreme wavelengths of the $K_s$ ($\Delta\lambda = 0.275\mu m$)
and $J$ ($\Delta\lambda = 0.290\mu m$) passbands to simulate the
diffusion of a wide-band source (Fig.\ref{ecran} down-right).
\end{itemize}
The contrast of the patterns are clearly considerably
affected by the size of the source (spatial coherence limitation),
but only marginally by the bandwidth (temporal coherence).

More details on this simulation will be published in a forthcoming paper
(\cite{Simul}).

\subsubsection{Modulation}
\begin{figure}[h]
\centering
\parbox{9.0cm}{
\includegraphics[width=8.cm, height=7.cm]{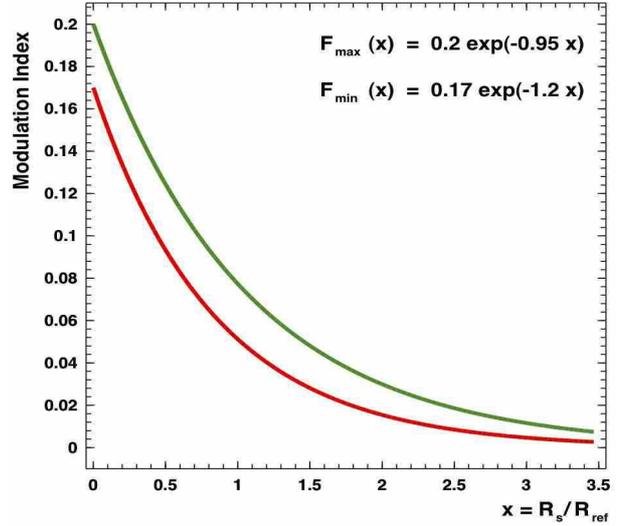}
}
\parbox{9.0cm}{
\caption[] 
{\it Expected
intensity modulation index $m_{scint.}=\sigma_I/\bar I$
of illumination patterns from simulated diffused stellar light
as a function of $x=R_S/R_{ref}$. The modulation indices are
essentially contained between the curves represented by functions
$F_{min}(x)$ and $F_{max}(x)$.
These functions allow one to constrain $x$ when the constraints
on $m_{scint.}$ are known.
}
\label{modindex}}
\end{figure}

In general, the small speckle from a point-source is almost
completely smoothed after the convolution by the projected stellar profile,
and only the large
structures of the size $R_{ref}(\lambda)$ -- or larger -- produce a significant modulation.
Therefore, as a result of the spatial coherence limitations,
the modulation of the illumination pattern critically depends on
the angular size of the stellar source  $\theta_s=r_s/(z_0+z_1)\sim r_s/z_1$.
Our Monte-Carlo studies 
show that the
intensity modulation index $m_{scint.}=\sigma_I/\bar I$
decreases when the ratio of the projected
stellar disk $R_S$  to the refraction scale $R_{ref}(\lambda)$ increases,
as shown in Fig. \ref{modindex}. This ratio can be expressed as
\begin{equation}
\frac{R_S}{R_{ref}(\lambda)}\! =\! \frac{r_s R_{diff}(\lambda)}{\lambda z_1}\! \sim \!
2.25\left[\frac{\lambda}{1\mu m}\right]^{-1}\!\!
\left[\frac{r_s/z_1}{R_{\odot}/10\,kpc}\right]
\left[\frac{R_{diff}(\lambda)}{1000\, km}\right]\! .
\label{contparam}
\end{equation}

At the first order, as intuitively expected,
the modulation index only depends on this ratio
$R_S/R_{ref}(\lambda)$ and not on other parameters of the phase screen
or explicitly on $\lambda$.
Indeed, the dispersion of $m_{scint.}$ for series of
{\it different} configurations generated with the same $R_S/R_{ref}$ ratio
is compatible with the statistical dispersion of $m_{scint.}$
in series of 10 patterns generated with {\it identical} configurations.
We empirically found that the modulation indices plotted as a function
of $R_S/R_{ref}$ are
essentially contained between the curves represented by functions
$F_{min}(x)=0.17 e^{-1.2 R_S/R_{ref}}$ and
$F_{max}(x)=0.2 e^{-0.95 R_S/R_{ref}}$ in Fig. \ref{modindex}.

\subsubsection{Time scale}
Because the 2D illumination pattern sweeps the Earth with a constant speed,
simulated light-curves of scintillating stars have been obtained by
regularly sampling the 2D illumination patterns along straight lines.
For a cloud moving with a transverse velocity $V_T$
with respect to the line of sight, the velocity
of the illumination pattern on
the Earth is $V_T (z_0+z_1)/z_1\sim V_T$ as $z_0\ll z_1$;
neglecting the inner cloud evolution
-- as in radio astronomy
(frozen screen approximation (\cite{lyne})) --
the flux variation at a given position is
caused by this translation, which induces
intensity fluctuations with a characteristic time scale of
\begin{eqnarray}
t_{ref}(\lambda)&=&\frac{R_{ref}(\lambda)}{V_T} \\
&\sim &
\!\! 5.2\, minutes\left[\frac{\lambda}{1\mu m}\right]\left[\frac{z_0}{1\, kpc}\right]\left[\frac{R_{diff}(\lambda)}{1000\, km}\right]^{-1}\left[\frac{V_T}{100\, km/s}\right]^{-1}\!\!\!\! . \nonumber
\end{eqnarray}
Therefore we expect the scintillation signal to be a stochastic fluctuation
of the light-curve with a frequency spectrum peaked around
$1/t_{ref}(\lambda)$ on the order of $(minutes)^{-1}$.

As an example, the scintillation at $\lambda=0.5\, \mu m$ of a LMC-star through
a Galactic $\mathrm{H_2-He}$ cloud located at $10\, kpc$ is characterized by
parameter
$R_S/R_{ref}(0.5\mu m)\sim (r_s/R_{\odot})\times(R_{diff}(0.5\mu m)/222\, km)$.
According to Fig. \ref{modindex},
one expects
$m_{scint.}=\sigma_I/\bar I>\,1\%$ if $x=R_S/R_{ref}\lesssim 2.4$
($=F_{min}^{-1}(0.01)$).
This will be the case for LMC (or SMC) stars smaller than the Sun as soon
as $R_{diff}(0.5\, \mu m)\lesssim 530\, km$.
If the transverse speed of the Galactic cloud is $V_T=200\, km/s$,
the characteristic scintillation time scale will be
$t_{ref}\gtrsim 24\, minutes$.

\subsection{Some specificities of the scintillation process}
In the subsections below we briefly describe the properties and
specificities of the scintillation signal that should be used to
distinguish a population of scintillating stars from the population
of ordinary variable objects.
\subsubsection{Chromaticity effect}
\label{sec:chromaticity}
Because $R_{ref}$ depends on $\lambda$, one expects
a variation of the characteristic time scale $t_{ref}(\lambda)$
between the red side of the optical
spectrum and the blue side. This property is probably one of
the best signatures of the scintillation, because
it points to a propagation effect, which is incompatible with
any type of intrinsic source variability.
\subsubsection{Relation between the stellar radius and the modulation index}
\label{sec:ssize}
As shown in Fig. \ref{modindex}, big stars scintillate less
than small stars through the same gaseous structure.
This characteristic
signs the limitations from the spatial coherence of the source and
can also be used to statistically distinguish the scintillating population
from other variable stars.
\subsubsection{Location}
\label{sec:location}
Because the line of sight of a
scintillating star has to pass through a gaseous structure,
we expect the probability for scintillation to be correlated with the
foreground --visible gas column-density. Regarding the invisible gas, it may
induce clusters of neighboring scintillating stars among
a spatially uniform stellar distribution
because of foreground --undetected gas structures.
This clustering without apparent cause
is not expected from other categories of variable stars.

\subsection{Foreground effects, background to the signal}
Conveniently, atmospheric {\it intensity} scintillation is negligible
through a large telescope ($m_{atm.}\ll 1\%$ for
a $>1\,$m diameter telescope~(\cite{dravins})).
Any other atmospheric effect such as absorption variations at
the minute scale (because of fast moving cirruses for example) should be easy
to recognize as long as nearby stars are monitored together.
Asteroseismology, granularity of the
stellar surface, spots or eruptions
produce variations of very different amplitudes and time scales.
A few types of rare recurrent variable stars exhibit emission
variations at the minute scale (\cite{sterken}),
but they could be identified from their spectrum or type.
Scintillation should also not be confused with absorption variations
caused by the dust distribution in the cloud;
indeed, the relative column density fluctuations
needed to produce measurable absorption variations ($\sim 1\%$) is higher
by several orders of magnitudes than the fluctuations that are able to produce
a significant scintillation (only a few $10^{-7}$ for
the clumpuscules, and $10^{-3}$ for the Bok globules
within a domain of $R_{diff}$ size).
\subsection{Expected optical depth}
\label{sec:optdepth}
Assuming a Galactic halo completely made of clumpuscules of
mass $M_c=10^{-3}M_{\odot}$, their sky coverage (geometrical
optical depth) toward the LMC
or the SMC should be on the order of $1\%$ according to
\cite{fractal1} (1994). This calculation agrees with a simple
estimate based on the density of the standard halo model taken from
(\cite{Caldwell}).
Here, we are considering only those structures that can be detected through
scintillation. Therefore we quantify the sky coverage of the
turbulent sub-structures that can produce this scintillation.
We define the {\it scintillation optical depth}
$\tau_{\lambda}(R_{diff.\ max.})$ as
the probability for a line of sight to cross a gaseous cloud with
a diffusion radius (at $\lambda$) $R_{diff}(\lambda)<R_{diff.\ max.}$;
this optical depth
is lower than (or equal to) the total sky coverage of the clumpuscules,
because it takes into account only those gaseous structures
with a minimum turbulence strength;
if $R_{diff} \rightarrow \infty$, all gaseous structures
account for the optical
depth and $\tau_{\lambda}(\infty)$ is the total sky coverage of the
clumpuscules.

\section{Feasibility studies with the NTT}
As shown in Fig. \ref{modindex},
the search for scintillation induced by transparent
Galactic molecular clouds makes it necessary to sample
at the sub-minute scale
the luminosity of LMC or SMC main sequence stars
with a photometric precision of -- or better -- than $\sim 1\%$.
In principle, this can be achieved with a two meter
class telescope with a high quantum efficiency detector
and a short dead-time between exposures.
To test the concept in a somewhat controlled situation,
we also decided to search for scintillation induced by known gas
through {\it visible} nebulae.

We found that the only setup
available for this short time-scale search was the ESO NTT-SOFI
combination in the infrared, with the additional benefit to enable the
monitoring of optically obscured stars that are located behind dark nebulae.
The drawback is that observations in infrared do not benefit from
the maximum of the stellar emission and the $3.6\, m$ diameter
telescope is barely sufficient to achieve the required photometric precision.

\subsection{The targets}
Considering the likely low optical depth of the scintillation process,
all our fields were selected to contain large numbers of target stars.
This criterion limited our search to the Galactic plane and to the LMC and SMC
fields.
For the search through visible nebulae (in the Galactic plane)
we added the following requirements:
\begin{itemize}
\item
Maximize the gas column density to benefit from a long phase delay.
Data from 2MASS (\cite{2MASS}) where used to select the clouds that
induce the stronger
reddening of background stars, pointing toward the thickest clouds.
\item
We selected nebulae that are strongly structured (from visual inspection)
and favored those with small spatial scale structures.
\item
We chose fields that contain a significant fraction of stars that are not
behind the nebula to be used as a control sample.
\end{itemize}
Our targets satisfy these requirements, except for B68, which does not match the
second criterion, but was selected so we could benefit from the large
number of published studies about this object.

We decided to observe toward the nebulae
with the $K_s$ filter to allow the monitoring of highly extincted stars
({\it i.e.} behind a high gas column density).

\begin{figure*}[!ht]
\begin{center}
\includegraphics[width=4.4cm]{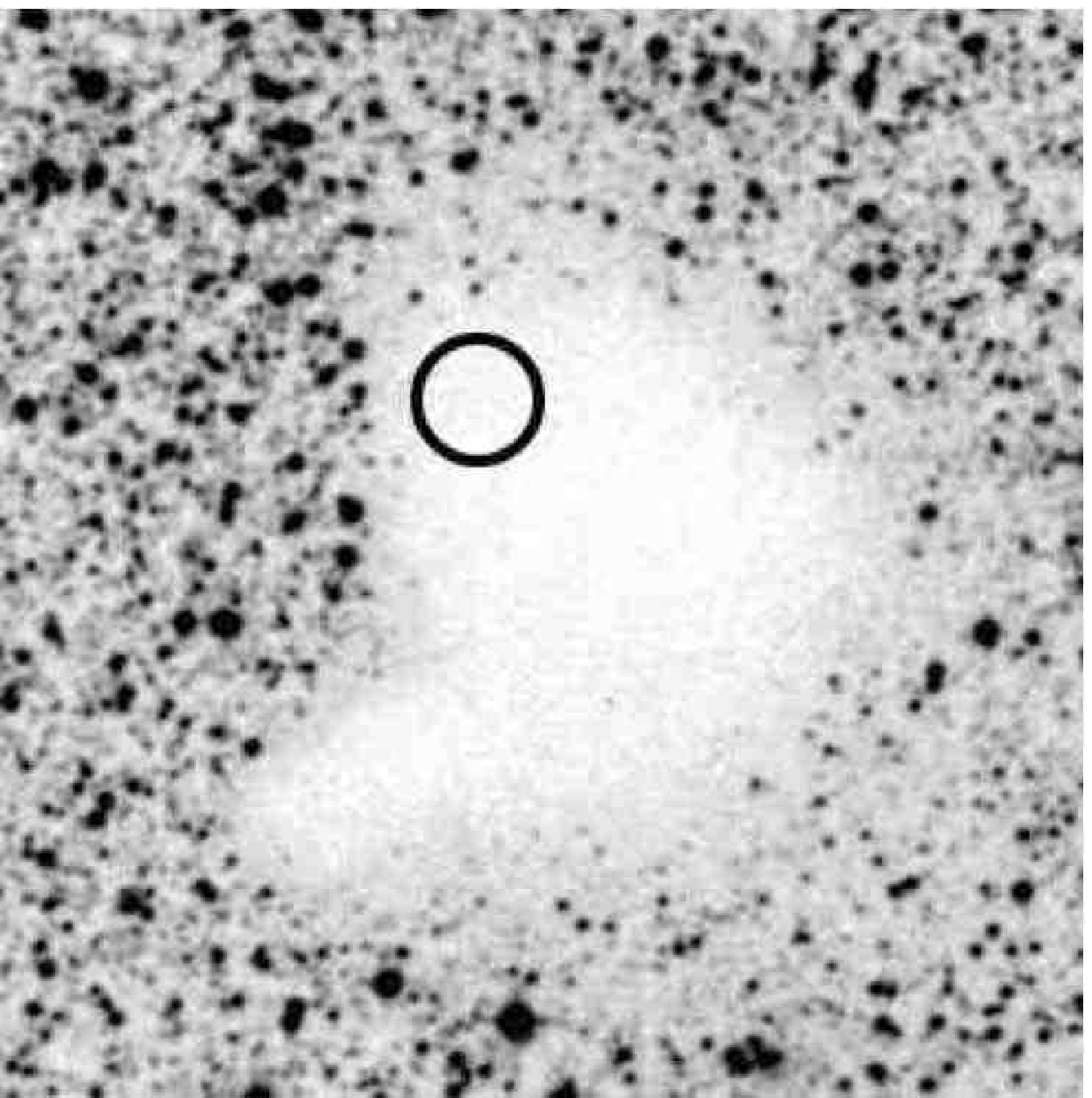}
\includegraphics[width=4.4cm]{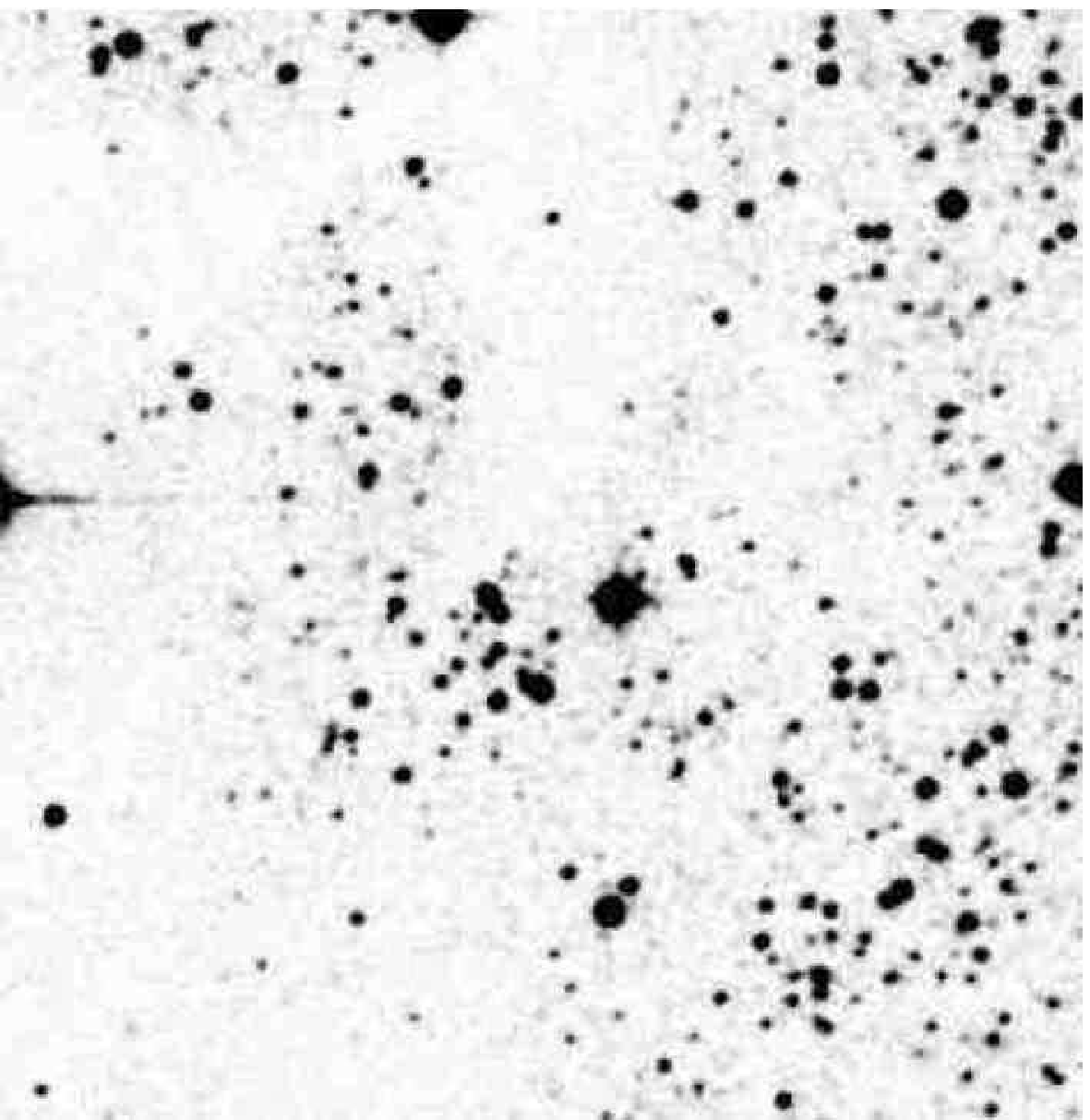}
\includegraphics[width=4.4cm]{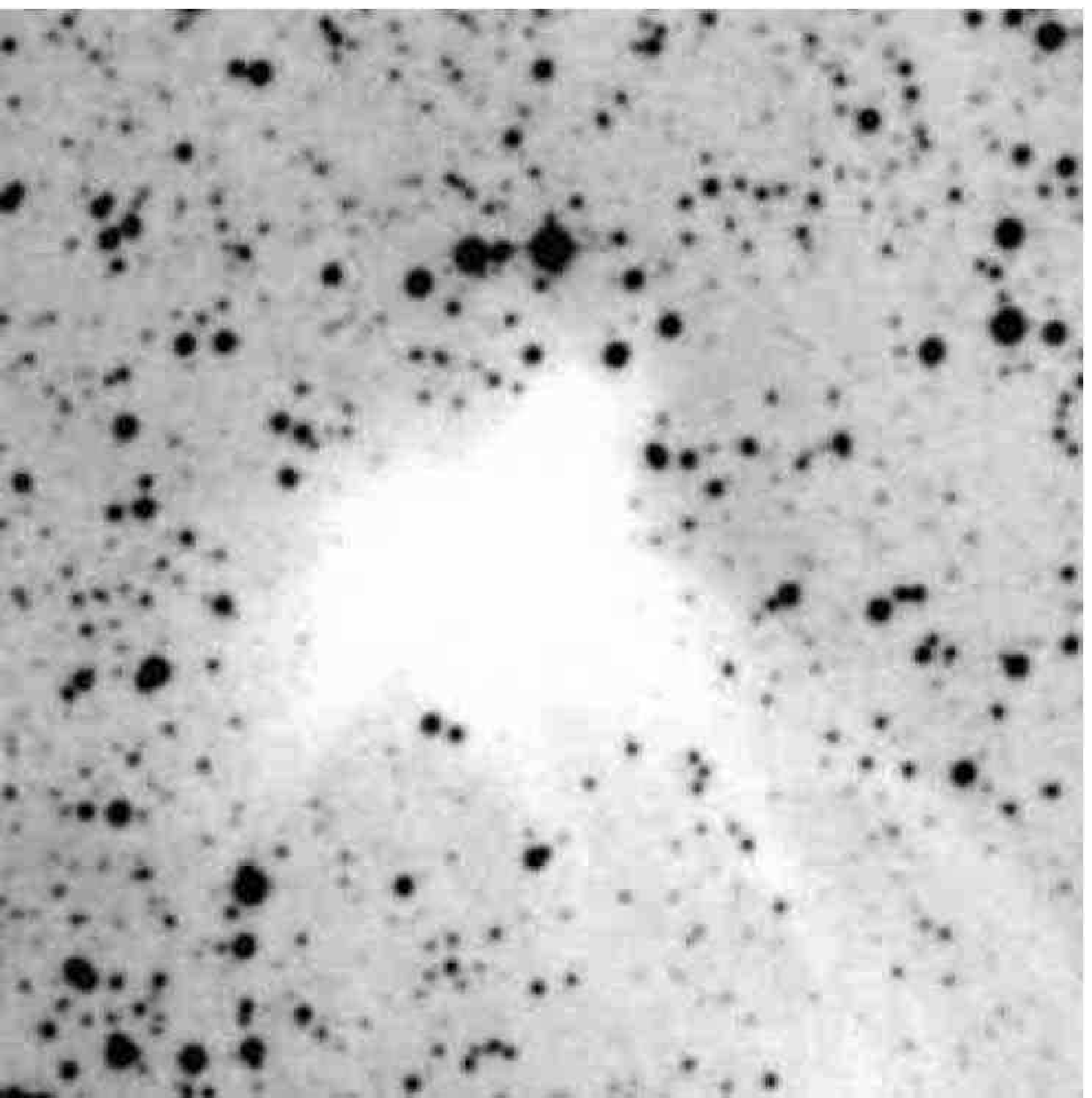}
\includegraphics[width=4.4cm]{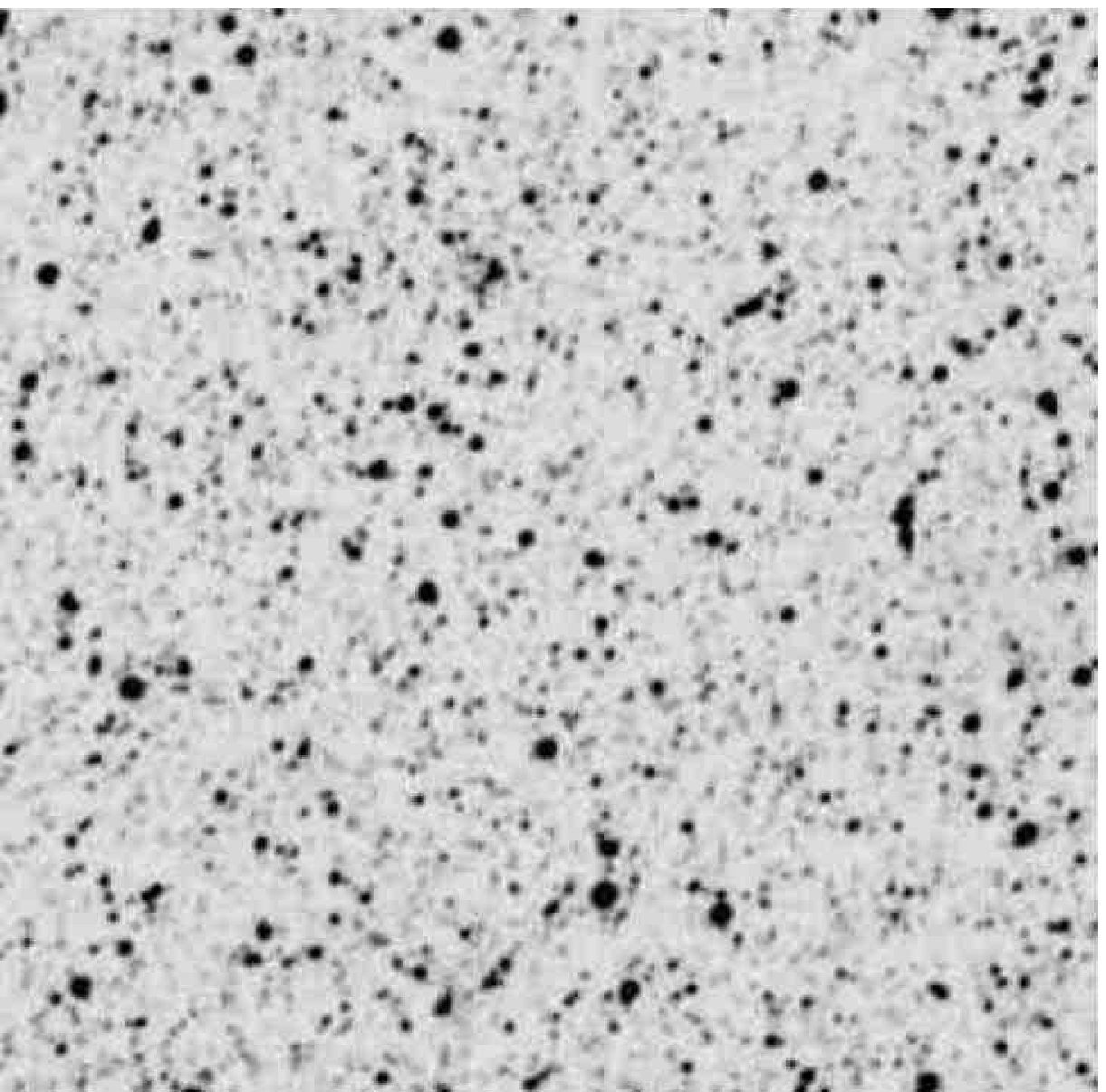}
\end{center}
\begin{center}
\includegraphics[width=4.4cm]{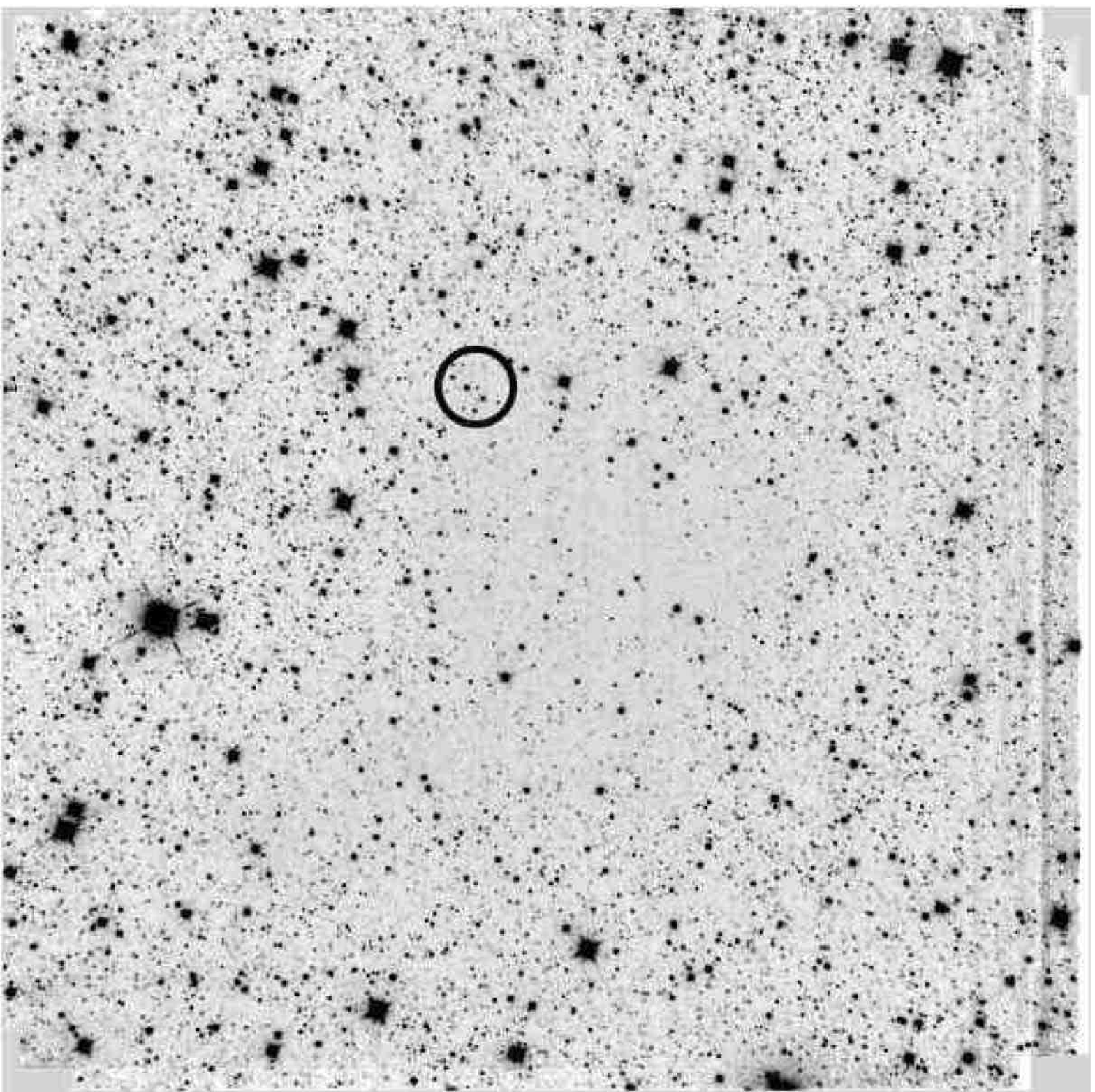}
\includegraphics[width=4.4cm]{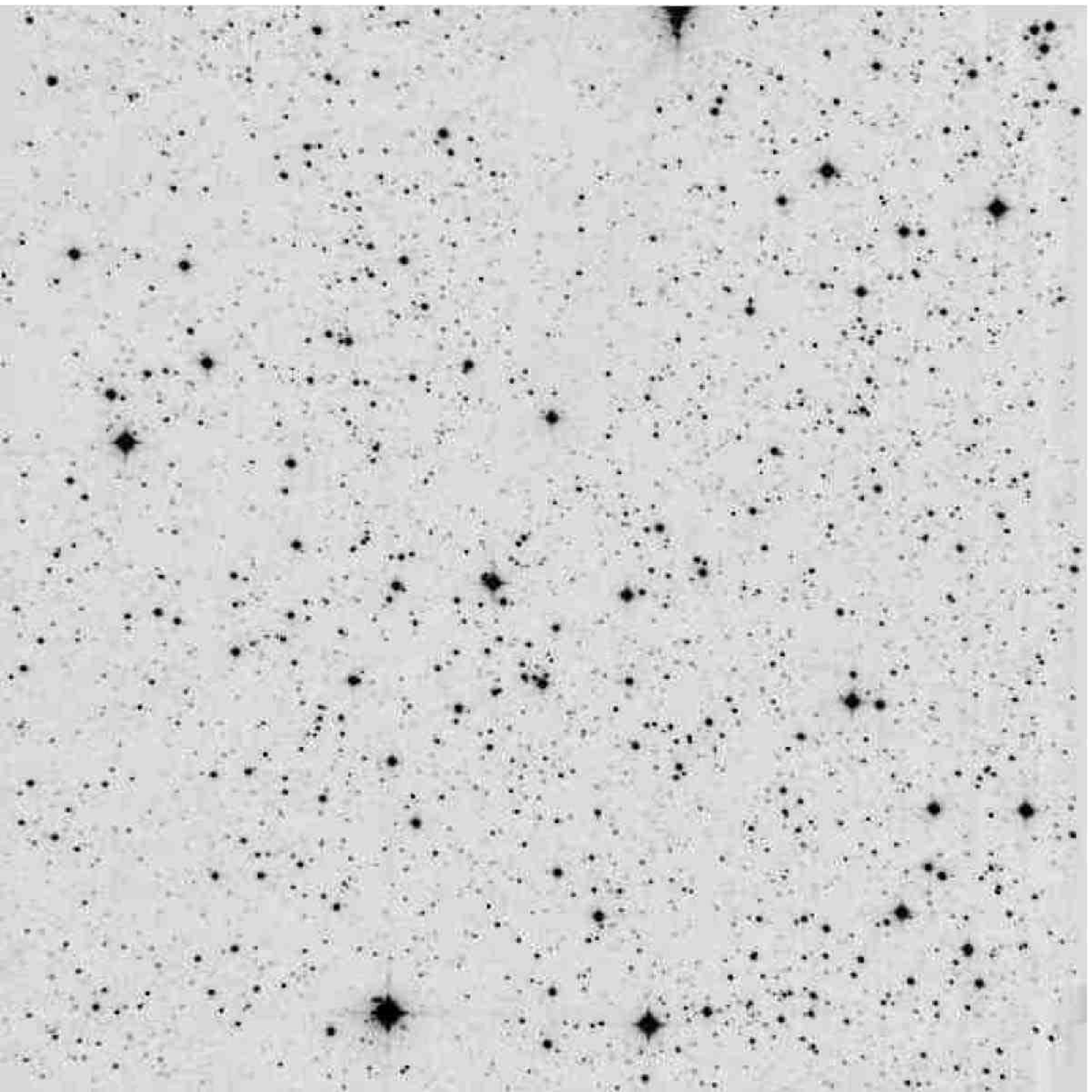}
\includegraphics[width=4.4cm]{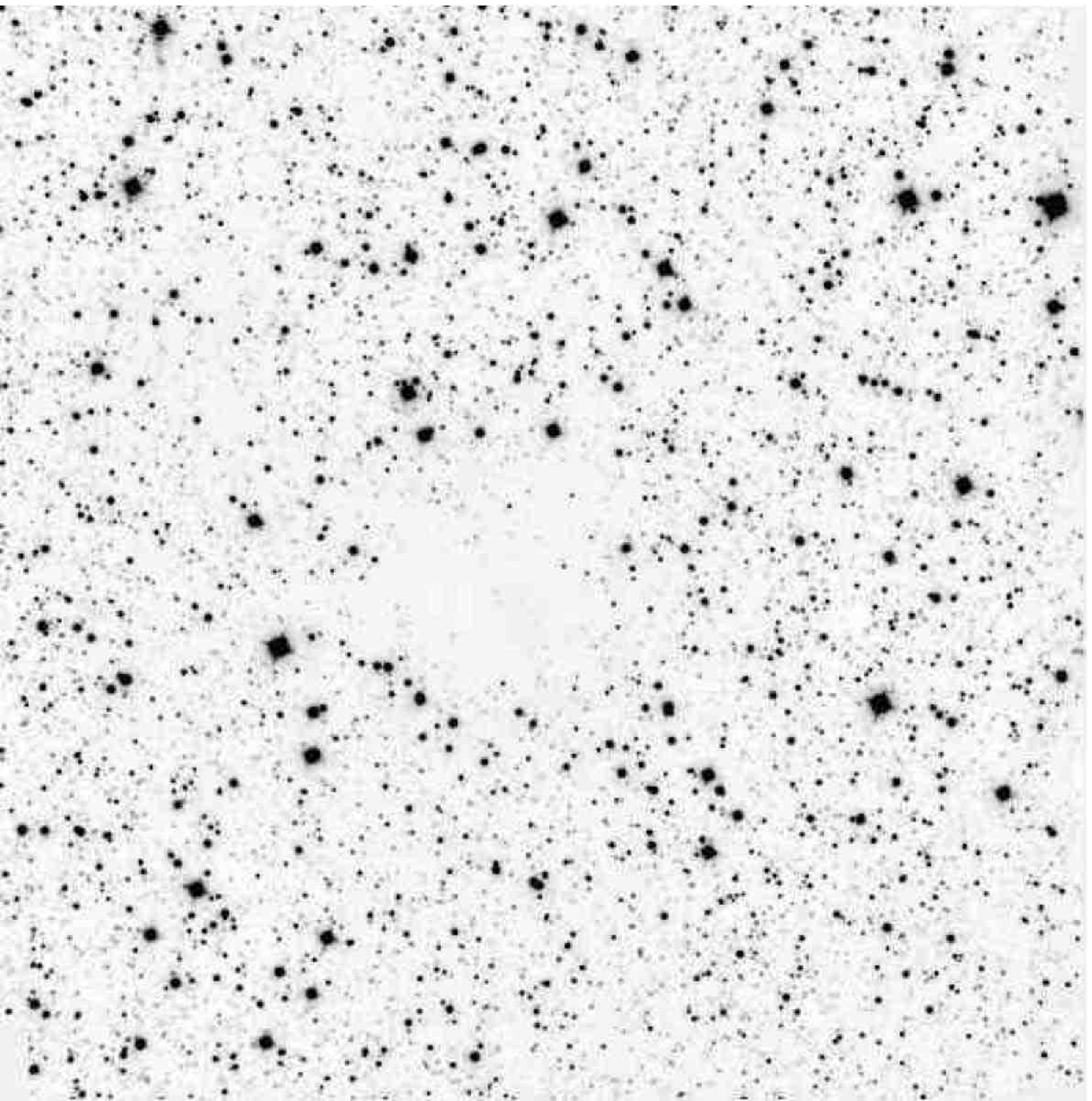}
\includegraphics[width=4.4cm]{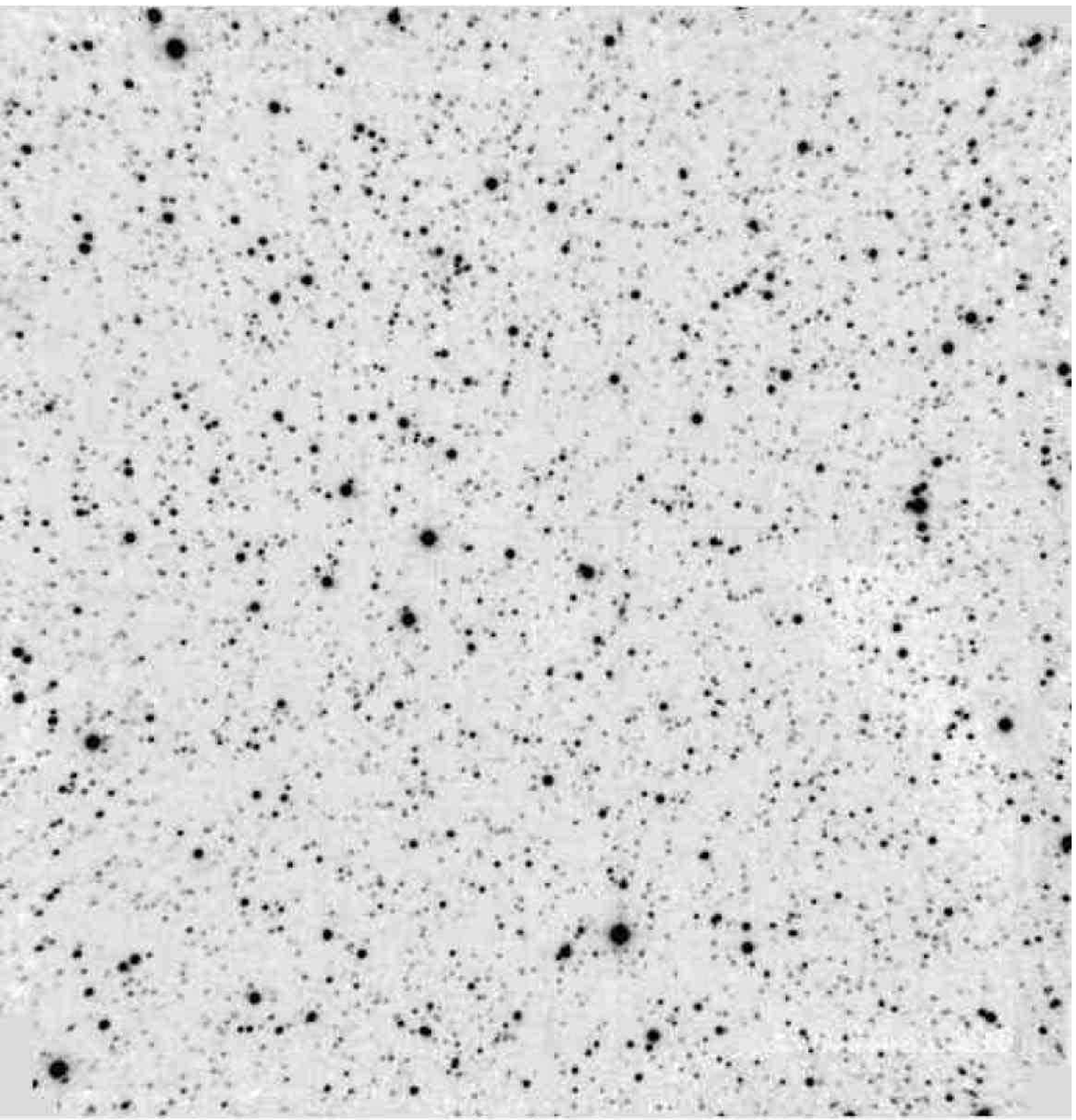}
\end{center}
\caption[] 
{\it
The four monitored fields, showing the structures
of the nebulae and the background stellar densities.
From left to right: B68, Circinus, cb131, and the SMC.
Up: images from the ESO-DSS2 in R. Down:
our corresponding template images (in $K_s$
for B68, cb131 and Circinus, in $J$
for SMC).
North is up and East is left.
The circles on the B68 images show the position of our
selected candidate.
}
\label{fields}
\end{figure*}

To make a test search for transparent (hidden) gas,
we selected a crowded field
in the SMC (LMC was not observable at the time of observations).
Then we used the $J$ filter to collect the maximum light fluxes
attainable with the SOFI detector.
Figure \ref{fields} shows the four monitored
$(4.92\times 4.92)\ Arcmin^2$ size fields in $R$ and in the $K_s$ or
$J$ passbands;
Table \ref{tab:targets} gives their list and characteristics
and also the main observational and analysis information.

\begin{table*}
\begin{center}
\begin{tabular}{c|c|c|c|c|}
			& SMC	& B68	& cb131	& Circinus \\
\hline
$\alpha$ (J2000)	& 00:52:41.3 & 17:22:40.7 & 18:16:59.4 & 14:59:28.9 \\
$\delta$ (J2000)	& -72:49:14.3 & -23:49:47.2 & -18.01:53.2 & -63:06:10.1 \\
central gas density		& - & $2.61\times 10^5 cm^{-3}$	& $1.8\times 10^5 cm^{-3}$ &	\\
center-to-edge density contrast & - & 16.5 & 140. &	\\
central column density Nl	& - & $2.59\times 10^{22} cm^{-2}$ & $5.8\times 10^{22} cm^{-2}$ &	\\
distance of nebula	& - & 80 pc		& 190 pc	& 170 pc\\
minor axis of nebula	& - & 17000 AU	& 24000 AU	& complex\\
distance of sources	& 62 kpc & $\sim 8\ kpc$	& $\sim 7\ kpc$	& $\sim 7\ kpc$	\\
\hline
light-curve duration night 1 (hours) 	& 2.18	& 4.77	& 1.25	& 1.84	\\
light-curve duration night 2 	& 2.63	& 5.07	& 1.74	& 2.16	\\
\hline
number of detected stars	& 5042	& 9599	& 9084	& 5249	\\
number of monitored stars	& 691	& 1114	& 2779	& 913	\\
magnitude of monitored stars	& $J<17.8$	& $K_s<17.1$	& $K_s<17.1$	& $K_s<17.1$	\\
fraction of stars behind dust	& $0\%$	& $46\%$& $64\%$ & -	\\
\hline
mean number of measurements/star & 980	& 2013	& 629	& 888	\\
\hline
\end{tabular}
\end{center}
\caption[]{\it
Observations and data reduction results. The data on the nebulae are taken
from \cite{Hotzel} (2002) for B68 and from \cite{Bacmann} (2000) for cb131.
The typical distance of the sources in the Galactic plane are taken
from \cite{georgelin} (1994) and \cite{russeil} (1998).
\label{tab:targets}
}
\end{table*}

Through the cores of B68 
and Circinus,
gas column densities of $\sim 10^{22}$ atoms/$cm^2$ induce an
average phase delay 
of $250\times 2\pi$ at $K_s$ central wavelength ($\lambda=2.16\,\mu m$).
According to our studies, a few percent scintillation signal is expected
from stars smaller than the Sun
if relative column density fluctuations of only $\sim 10^{-3}$ occur
within less than a few thousand kilometers
(corresponding to $R_{diff}\lesssim 2000\,$km).
These fluctuations -- which are probably rare -- could induce dust
absorption variations of only $\sim 10^{-3}$,
which can be neglected.
The expected time scale of the scintillation would be
$t_{ref}\gtrsim 5\, minutes$, assuming $V_T\sim 20\, km/s$.
\subsection{The observations}
During two nights of June 2006
we took a total of 4749 consecutive exposures of ${T_{exp}=10\,s}$,
with the infra-red $1024\times 1024$ pixel
SOFI detector in $K_s$ ($\lambda=2.16\,\mu m$) toward B68,
Circinus, cb131, and in $J$ ($\lambda=1.25\,\mu m$) toward the SMC
(see Table \ref{tab:targets} for details).
We recall that
for a dedicated search for transparent {\it hidden} matter,
measuring visible light ($B$, $V$, $R$ or $G$ filters)
-- corresponding to the maximum stellar emission -- would be better.
\section{Data reduction}
\subsection{Photometric reduction}
\begin{figure}[h]
\centering
\parbox{8cm}{
\includegraphics[width=8cm,height=6cm]{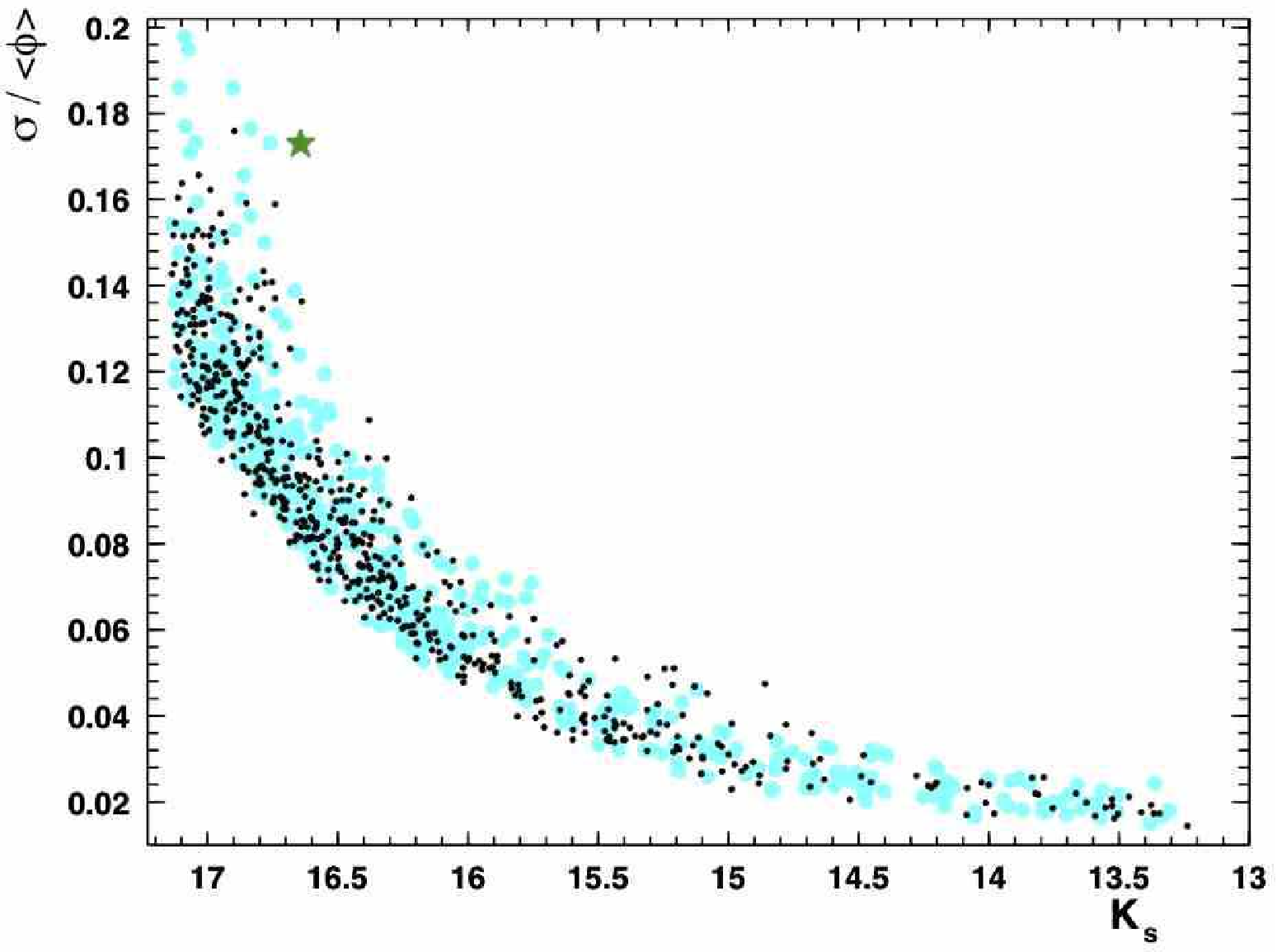}
\includegraphics[width=8cm,height=6cm]{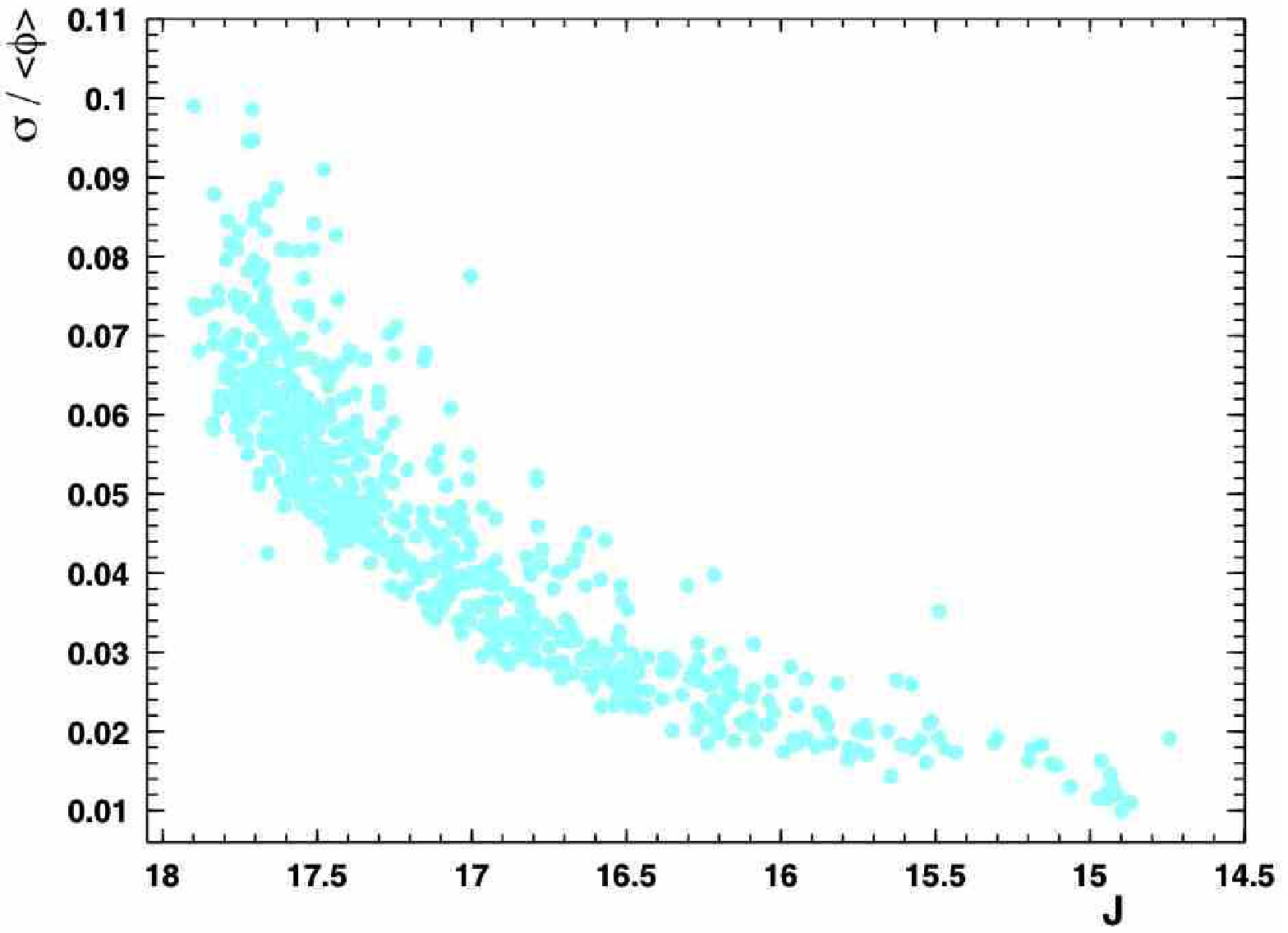}
}
\caption[]{\it
Dispersion of the photometric measurements along the light-curves
as a function of the mean $K_s$ magnitude (up) and $J$ magnitude (down).
Each dot corresponds to one light-curve.
In the upper panel, the small black dots correspond to control stars
that are not behind the gas; the big blue dots correspond to stars
located behind the gas; the big star marker indicates our selected candidate.
\label{resolution}}
\end{figure}

With the EROS software (\cite{PEIDA})
we produced the light-curves $\phi(t)$ of a few thousands of stars for
each target.
The reference catalogs of the monitored stars where established from
templates obtained through the standard GASGANO procedure (\cite{GASGANO})
by co-adding 10 exposures of 60s each
for the B68, Circinus, and cb131 fields (13 exposures for the SMC field).
The photometric measurements were all aligned with respect to
these templates.
We experimented with several techniques for the photometric reduction
(aperture photometry and Gaussian Point Spread Function (PSF) fitting).
We found that the most precise photometric technique, which provides the smallest
average point-to-point variation, is the Gaussian PSF fitting.
We observed that the Gaussian fit quality of the PSF decreases
when the seeing ({\it i.e.} the width of the PSF) is small.
This effect induces some correlation
between the seeing and the estimated flux, and we systematically
corrected the flux for this effect, according to a procedure
described in \cite{thesetisserand}.
Figure \ref{resolution} shows the dispersion of the
measurements along the light-curves.
This dispersion depends on the photometric precision
and on the -- possible but rare -- intrinsic stellar apparent variability.
We checked that it is not affected by the dust by considering
separately a control sample of stars (Fig. \ref{mask})
that are apart from the nebula (Fig. \ref{resolution}, upper panel).
We interpret the lower envelope of Fig. \ref{resolution} as the
best photometric precision achieved on (stable) stars at a given
magnitude. The outliers that are far above this envelope
can be caused by a degradation of the photometric precision in
a crowded environment, by the parasitic effect of bad
pixels, or by a real variation of the incoming flux.
The best photometric precision does not simply result from the Poissonian
fluctuations of the numbers of photoelectrons $N_{\gamma e}$. We
discuss the main sources of the precision limitation in Appendix B.
\subsection{Calibration}
The photometric calibration was done using the stars from the
2MASS catalog in our fields (\cite{2MASS}).
These stars were found only in the SMC
and B68 fields. Because all our fields were observed during the same nights
with stable atmospheric conditions, we extrapolated the calibration
in the $K_s$ band from B68 to cb131 and Circinus, after checking on series
of airmass-distributed images that the
airmass differences between the reference images (which are smaller
that 0.1) could not induce flux variations higher than 0.1 magnitude.
\section{Analysis}
\subsection{Filtering}
\begin{figure}[h]
\centering
\parbox{8.5cm}{
\includegraphics[width=4cm]{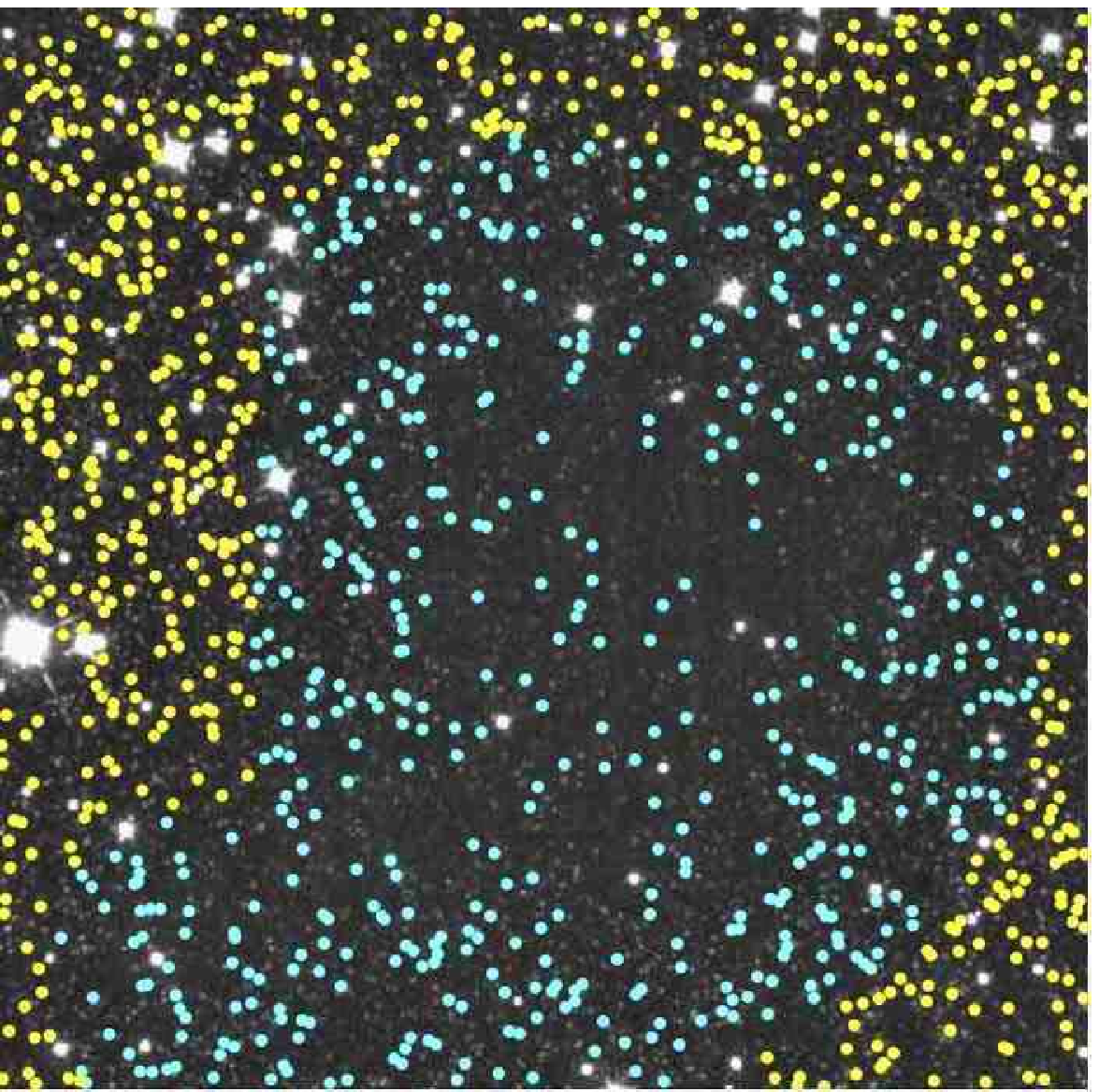}
\includegraphics[width=4cm]{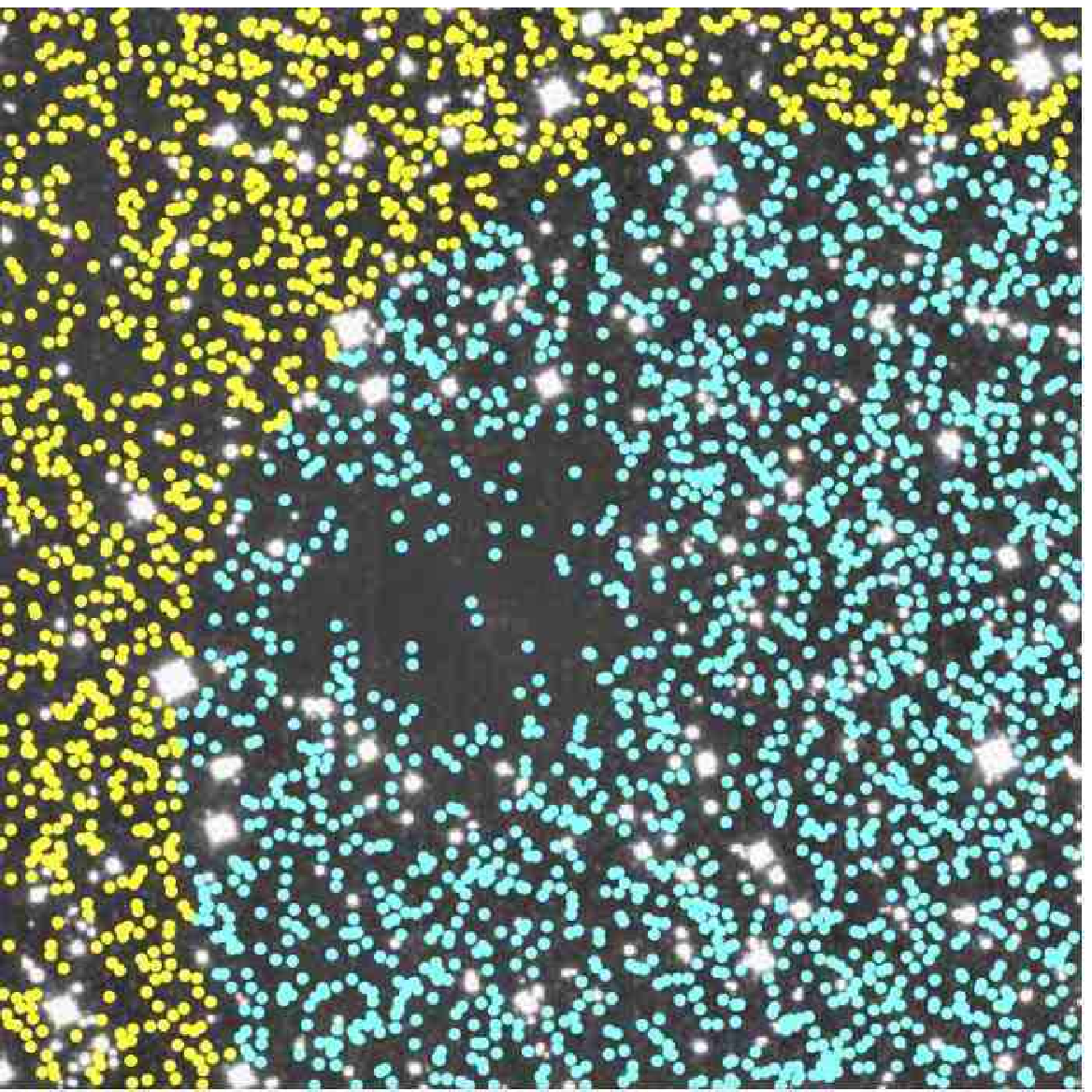}
}
\caption[]{\it
Definition of the control regions (yellow dots)
and the search regions (blue dots) toward B68 (left) and cb131 (right).
\label{mask}}
\end{figure}

We removed the lowest quality images, stars, and mearurements by requiring
the following criteria.
\begin{itemize}
\item
Fiducial cuts:
We do not measure the luminosity of objects closer than
100 pixels from the limits of the detector to avoid visible
defects on its borderlines.
The effective size of the monitored fields is then $4.44\times 4.44\ Arcmin^2$.
\item
We remove the images with poor reconstructed seeing by requiring
$s < \bar{s} + 2 \sigma_s$, where $\bar{s}$ and $\sigma_s$ are the
mean and r.m.s of the seeing distributions (see Fig. \ref{seeing}).
\begin{figure}[h]
\centering
\parbox{8cm}{
\includegraphics[width=8cm]{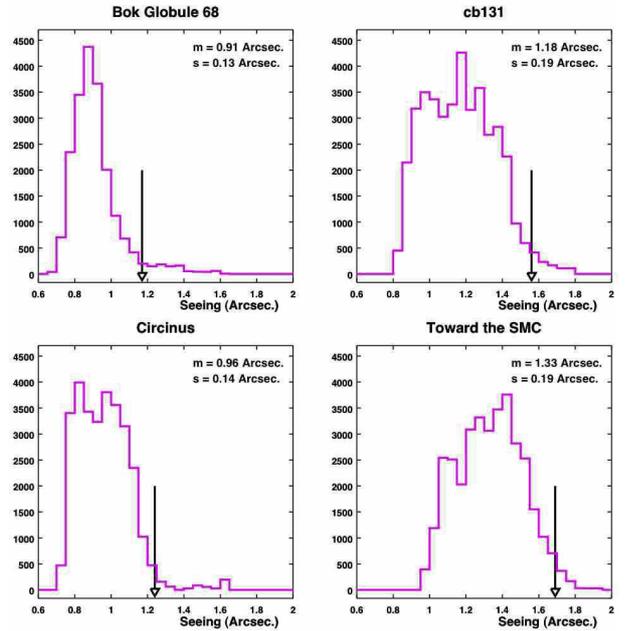}
}
\caption[]{\it
Seeing distributions of the images towards the four targets.
The images with larger seeing than the position of the arrow are discarded.
\label{seeing}}
\end{figure}
\item
We remove the images with an elongated PSF that have an eccentricity
$e>0.55$.
\item
We reject the measurements with a poor PSF fit quality.
This quality varies with the seeing, the filter, the position
of the object on the detector, and also with the flux of the
star. We require the $\chi^2$ of the fit to satisfy $\chi^2/d.o.f<4$.
\item
Minimum flux: We keep only stars with an instrumental flux
$\phi>1000\,ADU$ (corresponding to $J<17.8$, $K_s<17.1$)
to allow a reproducibility of the photometric
measurements better than $\sim 15\%$ in $K_s$ and $\sim 8\%$ in $J$.
The number of these stars are given in Table \ref{tab:targets}.
\item
We require at least 10 good quality measurements per night per light-curve
before searching for variabilities.
\end{itemize}
At this stage, we selected a set of
sufficiently sampled and well measured light-curves
to perform a systematic search for variabilities.
The numbers of light-curves we select for each target are given
in Table \ref{tab:targets}. As
already mentioned, we divided B68
and cb131 into two subfields containing control stars
(out of the nebula's field) and stars behind the nebula
(Fig. \ref{mask}).
Table \ref{tab:targets} gives the fraction of these stars located
behind visible dust, which traces the gas toward B68 and cb131.
The case of Circinus is different because we cannot clearly define a
control zone.

\subsection{Selecting the most variable light-curves}
\label{sec:selection}
\begin{figure}[h]
\centering
\parbox{8cm}{
\includegraphics[width=8cm]{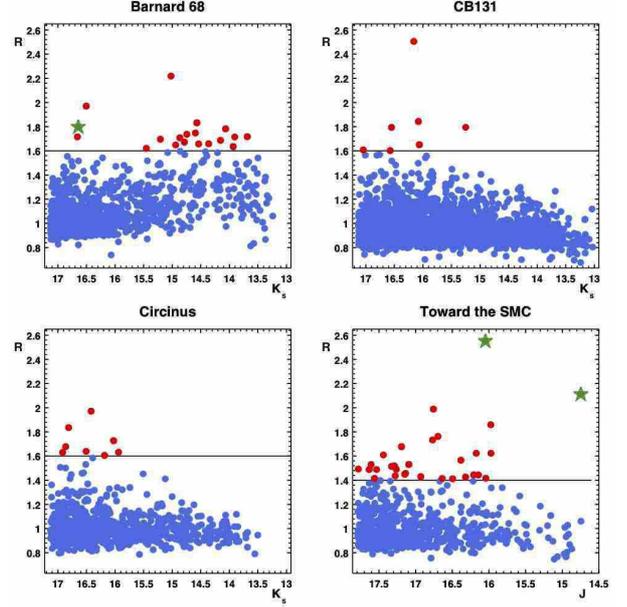}
}
\caption[]{\it
The $R=\sigma_{\phi}/\sigma_{int}$ versus $K_s$ (or $J$) distributions of the
light-curves of the monitored stars. The red dots correspond to
the most variable light-curves. The green stars indicate two known variable
objects toward SMC and our selected candidate toward B68.
\label{select}}
\end{figure}

We expect the scintillation signal to produce a stochastic fluctuation
of the incoming flux with a frequency spectrum peaked around
$1/t_{ref}(\lambda)$, on the order of $(minutes)^{-1}$.
The point-to-point variations of a stellar light-curve are caused by the
photometric uncertainties (statistical and systematical) and
by the possible intrinsic incoming flux fluctuations.
Note that fluctuations with time scales shorter than
the sampling ($\delta t \lesssim t_{i+1}-t_{i}\sim 15s$) are smoothed
in our data, and their high-frequency component cannot be detected.
To select a sub-sample that includes
the intrinsically most variable objects,
we used a simple criterion
-- not specific to a variability type --,
which allowed us
to both check our ability to detect known variable objects and explore new
time domains of variability.
Our criterion is based on the ratio $R=\sigma_{\phi}/\sigma_{int}$
of the light-curve dispersion $\sigma_{\phi}$ to the
``internal'' dispersion $\sigma_{int}$, defined as the r.m.s of
the differences
between the flux measurements and the interpolated values from the
previous and next measurements:
\begin{eqnarray}
&&\!\!\!\!\! \sigma_{int} = \\
&&\!\!\!\!\!\sqrt{\frac{1}{N_{meas.}}\sum_i \left[ \phi(t_i)\! -
\left(\phi(t_{i-1})+[\phi(t_{i+1})\! -\! \phi(t_{i-1})]
\frac{t_{i}-t_{i-1}}{t_{i+1}-t_{i-1}}\right)\right]^2 } , \nonumber
\end{eqnarray}
where $N_{meas.}$ is the number of flux measurements and
$\phi(t_i)$ is the measured flux at time $t_i$
(the typical $t_{i}-t_{i-1}$ interval is $\sim 15s$).
$\sigma_{int}$ quantifies the point-to-point fluctuations, whereas
$\sigma_{\phi}$ is the global dispersion of a light-curve.
$R$ is high as soon as there is a correlation between consecutive
fluctuations, either because of incoming flux variations or systematic
effects. 
Figure \ref{select} shows the distributions of $R$ versus
the apparent magnitude $K_s$ ($J$ for SMC).
By selecting light-curves with $R > 1.6$ (in $K_s$)
or $R > 1.4$ (in $J$) -- the red dots in Fig. \ref{select} --
we retain those with
a global variation over the two nights that is significantly
larger than the point-to-point variations.
Because we found that $R$ almost never exceeds the selection threshold in large
series of simulated uniform light-curves affected by Gaussian errors,
we systematically inspected each of the high $R$ objects.
We found that almost all of them are artifacts
(all the red dots in Fig. \ref{select});
we identified the following causes, which are related
to the observational conditions:
\begin{itemize}
\item
During the meridian transit of B68, the light-curves of
a series of bright stars located in two regions showed an abrupt flux
transition, correlated with the equilibrium change in the mechanics of
the telescope.
\item
Some stars looked temporarily brighter because of contamination
from the rotating egrets of bright stars.
\item
We also identified a few stars transiting near hot or dead pixels, which
induced a distorsion of the flux determination.
\end{itemize}
After elimination of these artifacts, only the light-curves marked with
a star marker in Fig. \ref{select} remain as reliable variable objects.
For this statistically limited test, this simple way of selecting variable
objects was sufficient, because we were able to visually inspect
each light-curve.
We also calculated the autocorrelation function of the light-curves
with no benefit with respect to this basic criterion because of the
artifact pollution.
For future high statitics observations, we plan to analyze the time
power spectra
of the light-curves, which should allow an automatic segregation of the
artifacts.

\subsection{Sensitivity of the analysis to the known variable objects}
\begin{figure}[h]
\centering
\parbox{9cm}{
\includegraphics[width=4.5cm]{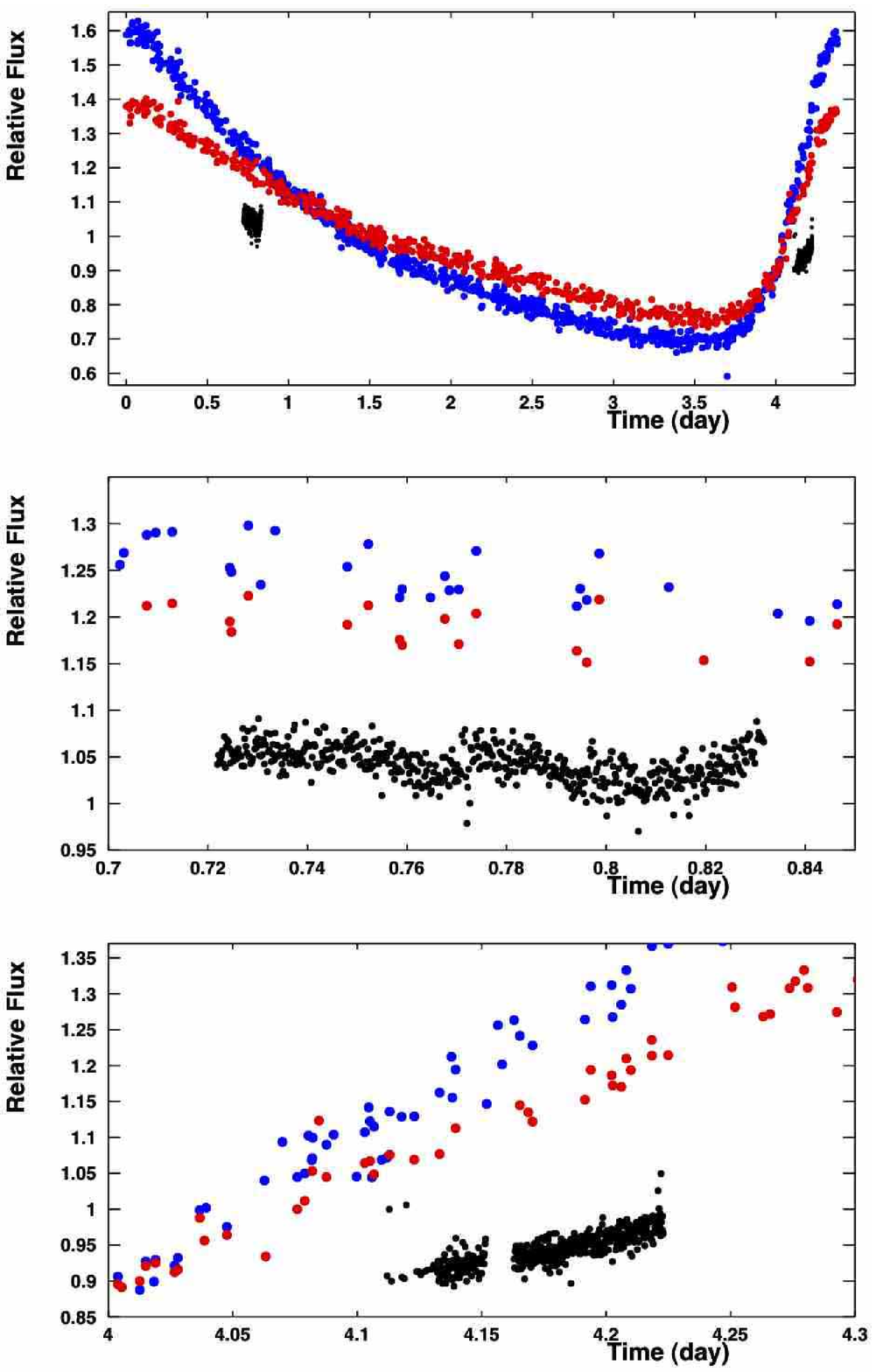}
\includegraphics[width=4.5cm]{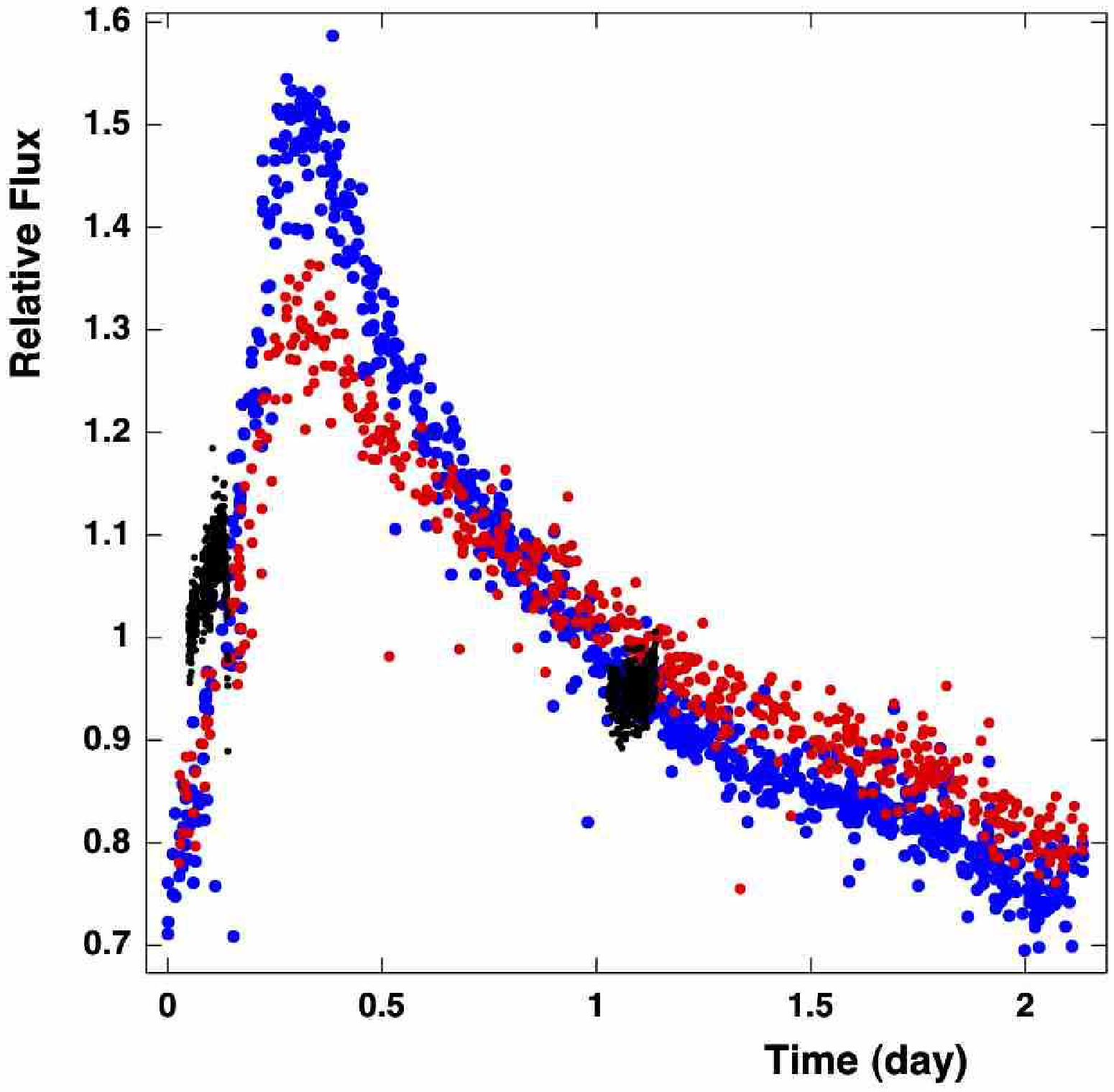}
}
\caption[]{\it
EROS folded light-curves of cepheids HV1562 (upper-left)
and of the EROS object ($13.2250\degree,-72.7951\degree$) (right)
in $B_{EROS}$ and
$R_{EROS}$ passbands, with our NTT observations in $J$ (black dots).
The lower-left panels show details around our observations.
\label{HV1562}}
\end{figure}

Our selection using variable $R=\sigma_{\phi}/\sigma_{int}$ is supposed
to retain any type of variability as soon as variations occur on a
time scale longer than our sampling time.
This allowed us to control our sensitivity to known variable stars.
This control has been possible only toward the SMC because it is the only field
where we found cataloged variable objects.
Indeed the CDS and EROS cataloges (\cite{ErosLMCfinal})
contain three variable objects within our SMC-field.
We
were able to identify two cepheids,
$HV1562$
($\alpha =13.1550\degree,\delta =-72.8272\degree,\ J2000$, $J=14.7$,
periodicity 4.3882 days)
and the EROS cepheid
($\alpha =13.2250\degree,\delta =-72.7951\degree,\ J2000$, $J=16.03$,
periodicity 2.13581 days).
Figure \ref{HV1562} shows the corresponding folded
light-curves from the EROS SMC database (phase diagram)
(\cite{ErosLMCfinal}, \cite{Hamadache}, \cite{Rahal}),
on which we superimpose our own observations.

We are content to note that
the light-curves of both objects were successfully selected
by our analysis; they correspond to the objects marked by a star
in Fig. \ref{select} (SMC).
Our precision was sufficient to clearly observe
the rapidly ascending phase of HV1562 during night 2,
as can be seen in Fig. \ref{HV1562}.

The third variable object in the SMC field is OGLE SMC-SC6 148139
($\alpha= 13.1446\degree,\delta= -72.8333\degree,\ J2000$, $B=16.5$).
It is a detached eclipsing binary with a periodicity of 1.88508 days.
Because no eclipse occured during the data taking, this object was
-- logically -- not selected as a variable by our analysis.

\subsection{Signal?}
\begin{figure}[h]
\centering
\parbox{8cm}{
\includegraphics[width=8cm]{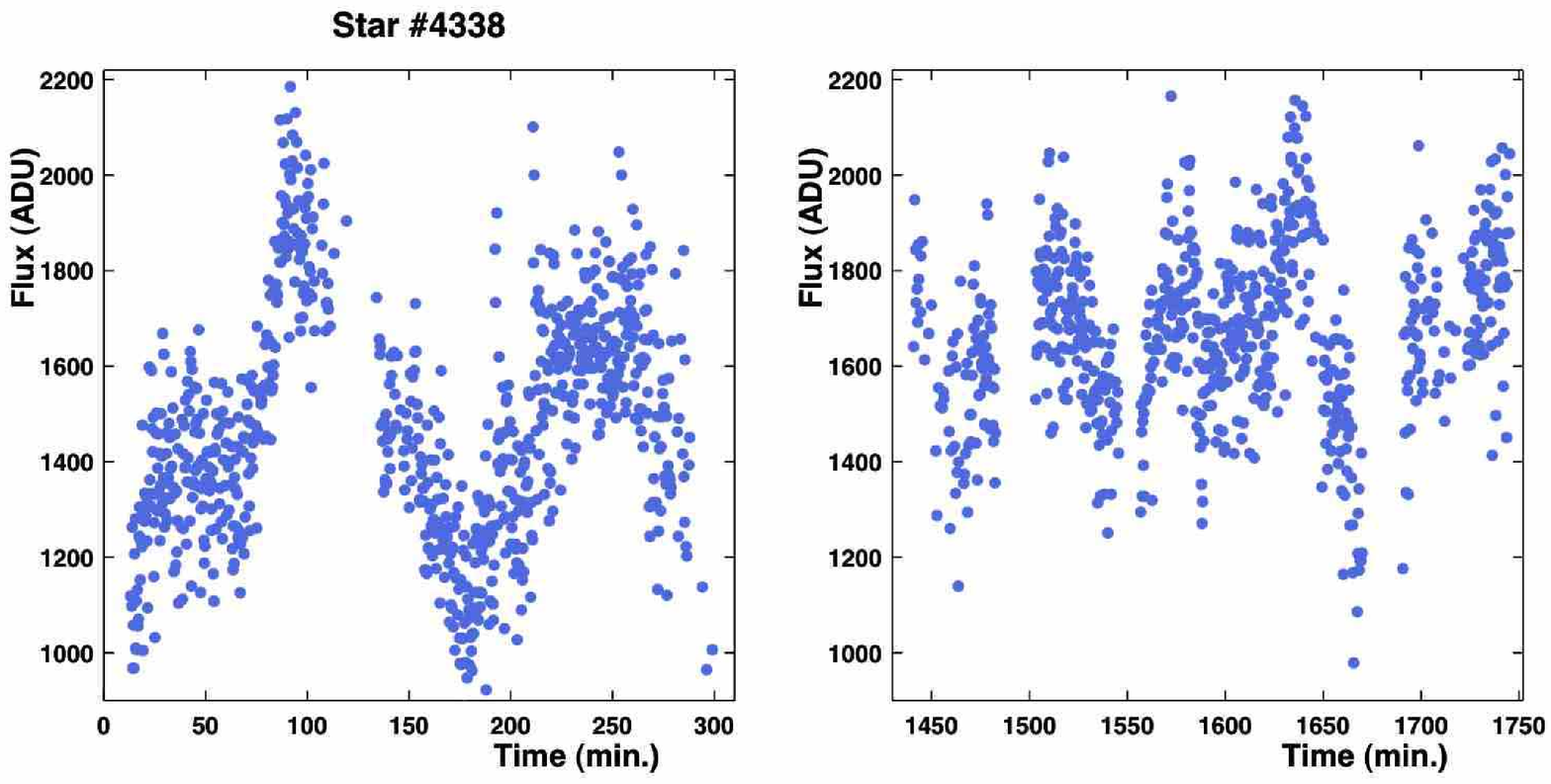}
\includegraphics[width=7cm]{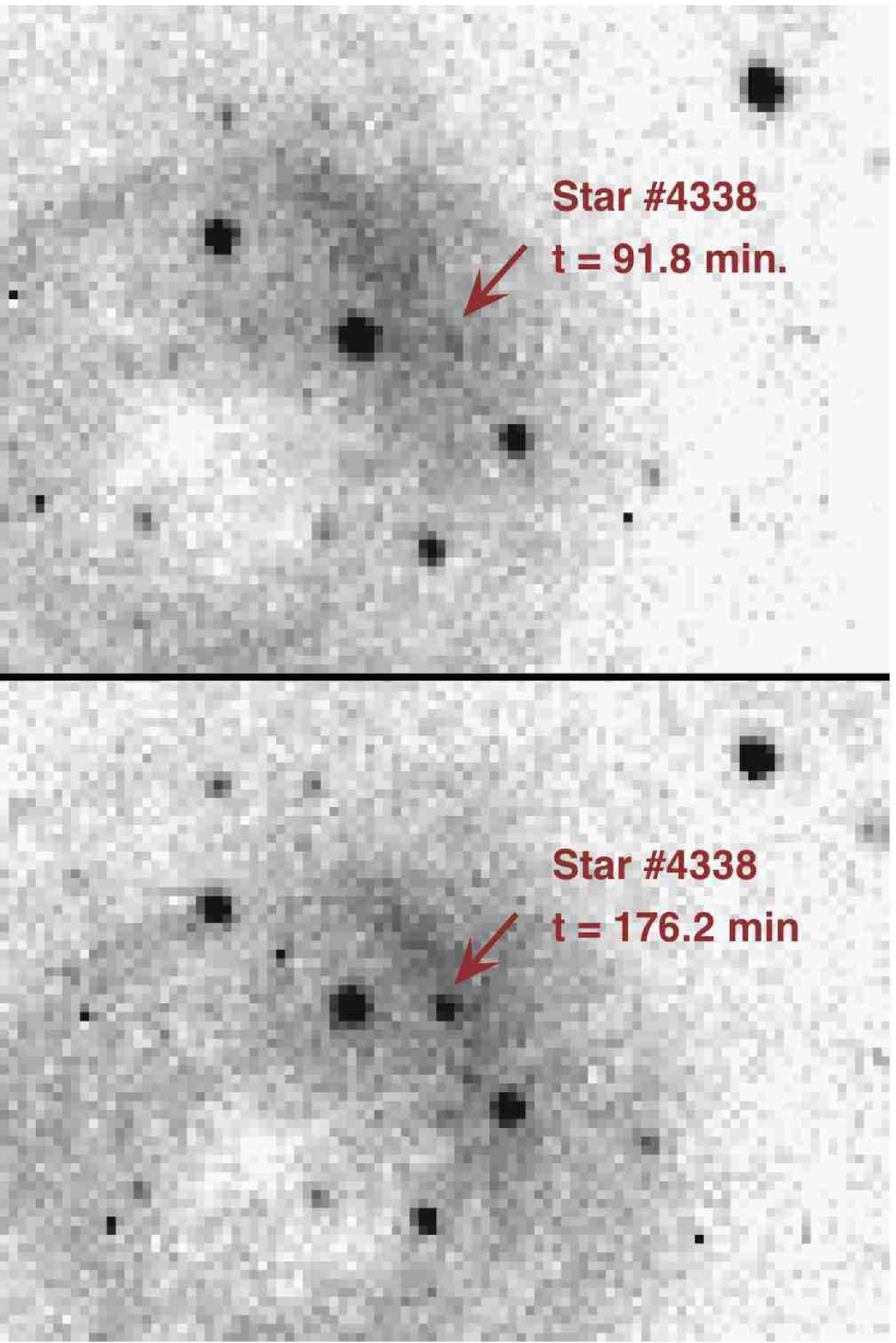}
}
\caption[]{\it
Light-curves for the two nights of observation (top) and images of the
selected candidate toward B68 during a low-luminosity phase (middle)
and a high-luminosity phase (bottom); North is up, East is left.
\label{candidate}}
\end{figure}

Only one variable object remains toward B68 after removing all artifacts
(see Fig. \ref{candidate} and see its location
$\alpha =260,6762\degree,\delta =-23.8159\degree\ (J2000)$
in Fig. \ref{fields}-left).
The star has magnitudes $K_s=16.6$ and $J=20.4$; its light is absorbed
by dust by $A_K=0.99$ magnitude; it is a main sequence star with
possible type ranging from A0 at $9.6\, kpc$ ($r_s=2.4\, R_{\odot}$) to
A5 at $6.1\, kpc$ ($r_s=1.7\, R_{\odot}$) or from F0 at $5.0\, kpc$
($r_s=1.6\, R_{\odot}$) to F5 at $4.0\, kpc$ ($r_s=1.4\, R_{\odot}$).
Consequently, it is a star small enough to experience observable scintillation.
The modulation index of the light-curve is $m=0.17$, which is
quite high but not incompatible with a variation owing to scintillation.
If we consider that this modulation is caused by a scintillation effect,
then it is
induced by a turbulent structure with $R_S/R_{ref}<0.25$, according
to Fig. \ref{modindex}; as we know that $r_s/z_1 > 2.5 R_{\odot}/10\,kpc$,
we directly conclude from expression (\ref{contparam})
that $R_{diff}(2.16\mu m)<96\, km$;
using expression (\ref{expression-rdiff}),
this gives the following constraint on the
size and the density fluctuations of the hypothetic turbulent structure:
\begin{equation}
\sigma_{3n}>1.45\times 10^9 cm^{-3} \left[\frac{L_z}{17000\, AU}\right]^{-\frac{1}{6}}.
\end{equation}
The largest possible outer scale $L_z$ corresponds here to
the minor axis of B68 ($17000\, AU$), but smaller turbulent structures within the
global system may also happen.

A definitive conclusion on this candidate would need complementary
multi-epoch and
multicolor observations, as explained in Sect. \ref{sec:chromaticity};
but the hypothetic
turbulent structure that is possibly responsible for scintillation has
probably moved from the line of sight since the time of observations,
considering its typical size.
Nevertheless, reobserving this object would allow one to check for
any other type of variability.
Considering the short time scale and the large amplitude
fluctuations, a flaring or eruptive star may be suspected, but probably not a
spotted star or an effect of astroseismology.

An important result comes from the rarity of these fluctuating objects:
There is no significant population of variable stars that can
mimic scintillation effects, and future searches 
should not be overwhelmed by background of fakes.

Ideally, for future observation programs aiming for an unambiguous signature,
complementary multiband observations should be planned shortly after the
detection of a scintillation candidate.
After detection of several candidates, one should also
further investigate the correlation between
the modulation index and the estimated gas column density to reinforce the
scintillation case (see Sect. \ref{sec:ssize} and \ref{sec:location}).

\section{Establishing limits on the diffusion radius}
In the following sections we will establish upper limits on the
existence of turbulent gas bubbles based on the observed
light-curve modulations.
The general technique consists to find
the minimum diffusion radius $R_{diff}$ 
for each monitored star that is compatible
with the observed modulation.

{\bf The smallest diffusion radius compatible with observed stellar light-curve fluctuations.}\\
Let us consider a star with radius $r_s$, placed at distance $z_1$ behind
a screen located at $z_0$ (with a projected radius $R_S=r_s\times z_0/z_1$).
If a turbulent structure characterized by $R_{diff}$ (with the corresponding
$R_{ref}$ given by Eq. (\ref{Rref}))
induces scintillation of the light of this star,
the corresponding modulation $m_{scint.}$ is within the limits
\begin{equation}
F_{min}(R_S/R_{ref})<m_{scint.}<F_{max}(R_S/R_{ref}), \label{limitm}
\end{equation}
as predicted from Fig. \ref{modindex}.

Below we assume that our time sampling is sufficient
to take into account any real variation within a time scale longer than
a minute.
Then, because the observed modulation $m$ of a stellar light-curve results from the
photometric uncertainties and from the hypothetic intensity
modulation $m_{scint.}$, one can infer that $m_{scint.}\le m$.
This inequality combined with inequality (\ref{limitm}) leads to
$F_{min}(R_S/R_{ref})< m$. Because $F_{min}$ is a decreasing function,
it follows that $R_S/R_{ref} > F_{min}^{-1}(m)$.
Using Eq. (\ref{contparam}) and
inverting function $F_{min}(x)=0.17 e^{-1.2 x}$,
this can be expressed
as a constraint on the value of $R_{diff}$ for the
gas crossed by the light:
\begin{equation}
R_{diff} > R_{diff}^{min} \sim 370\, km \left[\frac{\lambda}{1\mu m}\right]
\left[\frac{r_s/z_1}{R_{\odot}/10\,kpc}\right]^{-1}
\ln\left[\frac{0.17}{m}\right].
\label{rdiffmin}
\end{equation}

{\bf Information on the source size.}

To achieve a star-by-star estimate of $R_{diff}^{min}$,
it appears that
we need to know $z_1$, the distance from the screen to each source, and
each source size $r_s$. 
But because $z_1 >> z_0$, it follows that $z_1\sim z_0+z_1$,
and therefore the knowledge of the
angular stellar radius $\theta_s=r_s/(z_0+z_1)\sim r_s/z_1$
-- which can be extracted from the apparent magnitude and
the stellar type -- is sufficient for this estimate.

We can extract constraints on the stellar apparent radius $\theta_s$
from the $K_s$ (or $J$) apparent magnitudes by using
the following relation derived from the standard Stefan-Boltzman
blackbody law:
\begin{eqnarray}
\label{rvsk}
&&\!\!\!\!\!\log\left[\frac{\theta_s}{\theta(R_{\odot}\ at\ 10 kpc)}\right] =  \\
&&\!\!\!\!\!3-\frac{K_s\,(or\, J)}{5}-\frac{(V\! -\! K_s\, or\, J)}{5}+
\frac{M_{V\odot}\! +\! BC_{\odot}-BC}{5}-
2\log\left[\frac{T}{T_{\odot}}\right], \nonumber
\end{eqnarray}
where $R_{\odot}$ $M_{V\odot}$, $BC_{\odot}$, $T_{\odot}$ are
the solar radius, absolute V magnitude, bolometric correction and temperature;
$\theta(R_{\odot}\ at\ 10 kpc)$ is the angular solar radius at $10 kpc$;
$(V-K_s\, or\, J)$ taken from (\cite{Johnson}), $BC$ and $T$
taken from (\cite{Allen}) are the color index,
bolometric correction, and temperature respectively that characterize the type
of the star (independently of its distance).
We use this relation to establish the connection between $K_s$ (or $J$)
and the angular stellar radius $\theta_s$ for a given stellar type.
If the distance to the star is known,
then its type is directly obtained from its location in the color-magnitude
diagram;
if the distance is uncertain, spanning a given domain, we estimate
an upper value of $\theta_s$ as follows:
for each stellar type, we calculate the distance where the
apparent magnitude of the star would be $K_s$. If this distance is
within the allowed domain, we estimate $\theta_s$ from Eq. (\ref{rvsk}).
If the branch of the star (main sequence or red giant) is known, we
restrain the list of types accordingly. We conservatively consider
the maximum value $\theta_s^{max}$ found with this procedure.
In the next two sections, we will use Eq. (\ref{rdiffmin})
together with the stellar size constraints for each monitored star
to establish distributions of $R_{diff}^{min}$
toward the SMC and toward the dark nebulae.
\section{Limits on turbulent Galactic hidden gas toward the SMC}
\begin{figure}[h]
\centering
\parbox{9cm}{
\includegraphics[width=4.5cm]{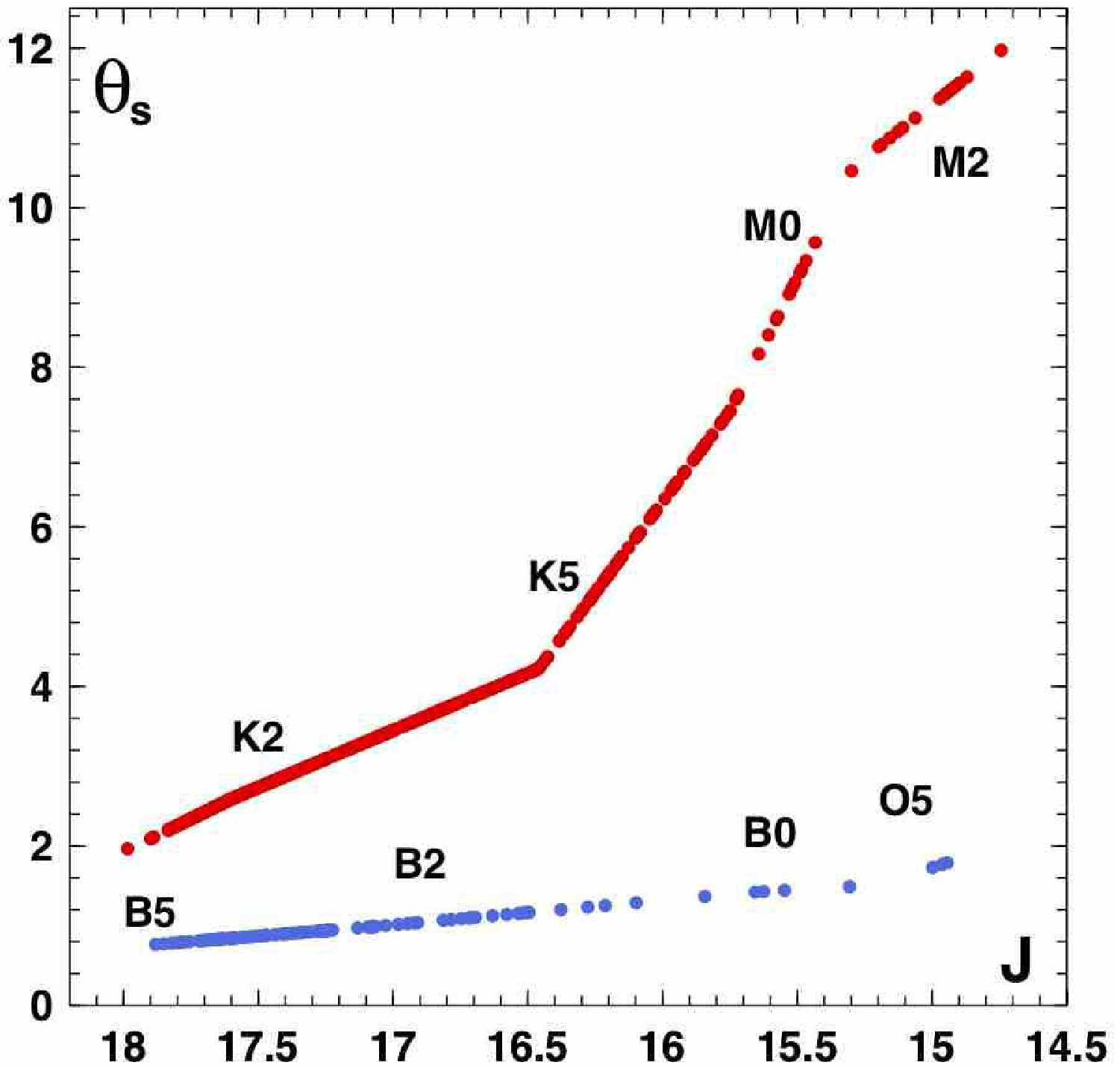}
\includegraphics[width=4.5cm]{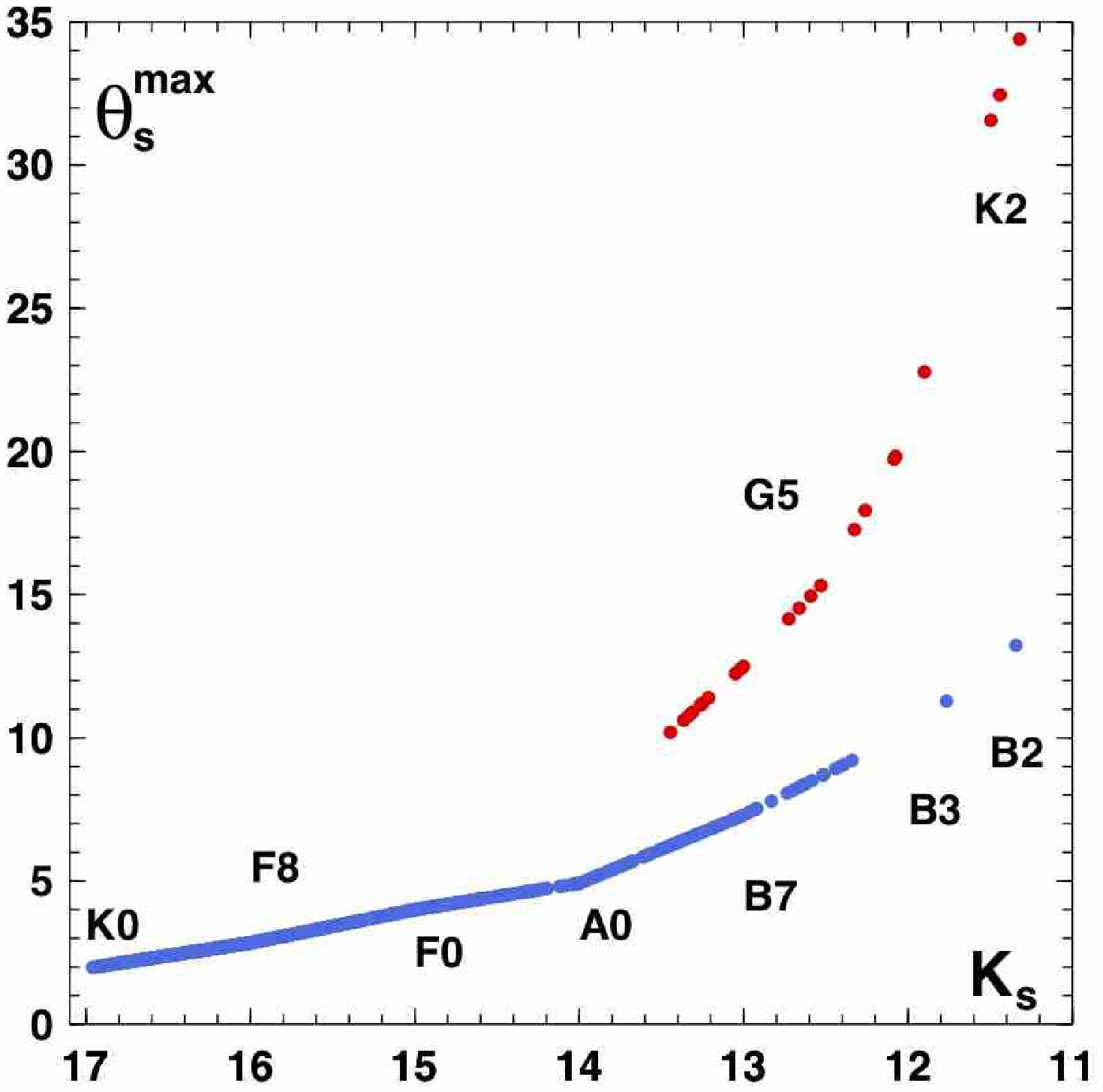}
}
\caption[]{\it
(left) Angular radius $\theta_s$ and type of the SMC stars as a function
of their apparent magnitude $J$ (in units of
the angular radius of the Sun at $10 kpc$). \\
(right) Maximum angular radius $\theta_s^{max}$ and corresponding type
of the B68
stars as a function of the apparent magnitude $K_s$.
The blue lines belong to the main sequence stars, the red lines
belong to the red giant stars.
\label{thetas}}
\end{figure}

\begin{figure}[h]
\centering
\parbox{8cm}{
\includegraphics[width=8cm,height=6cm]{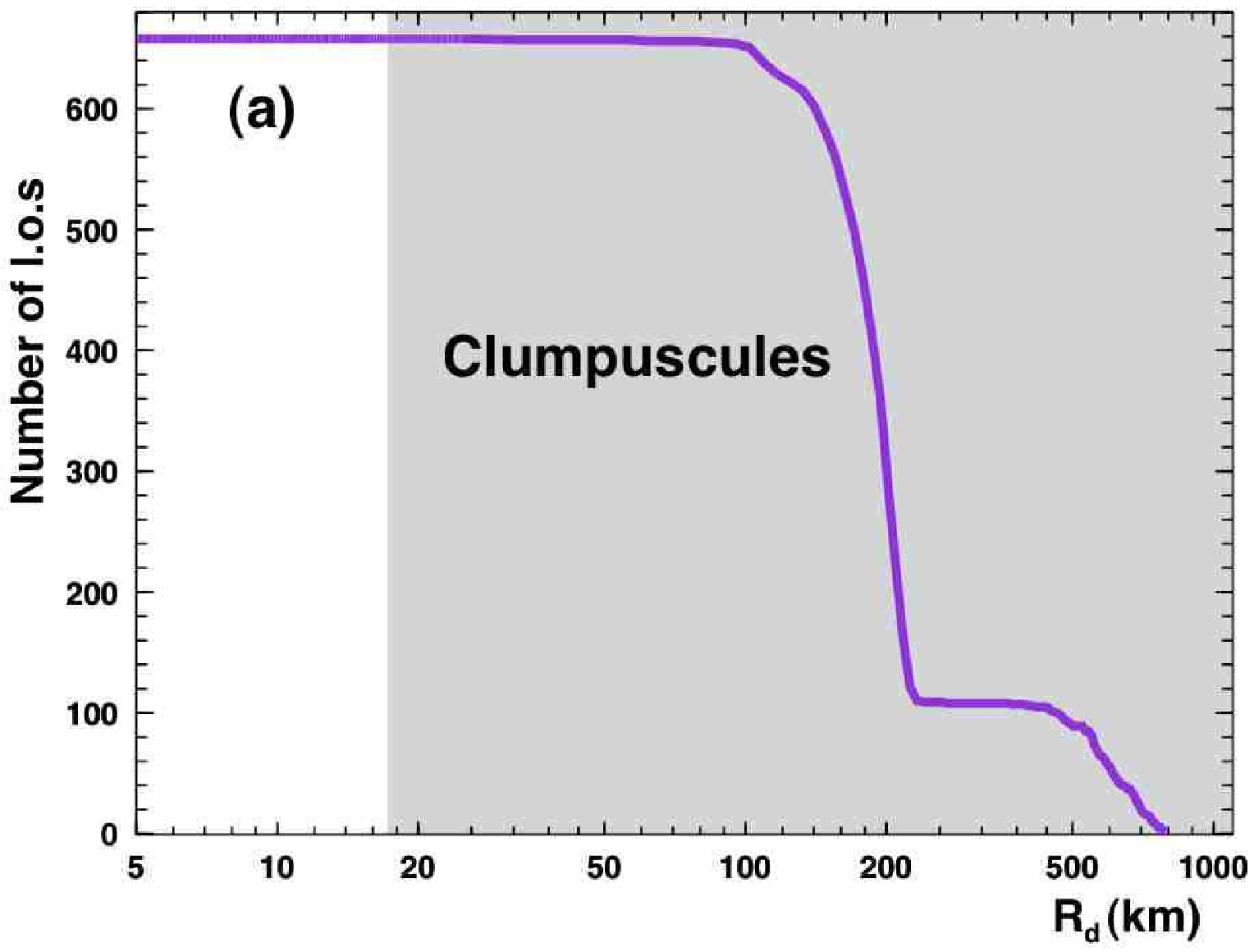}
\includegraphics[width=8cm,height=6cm]{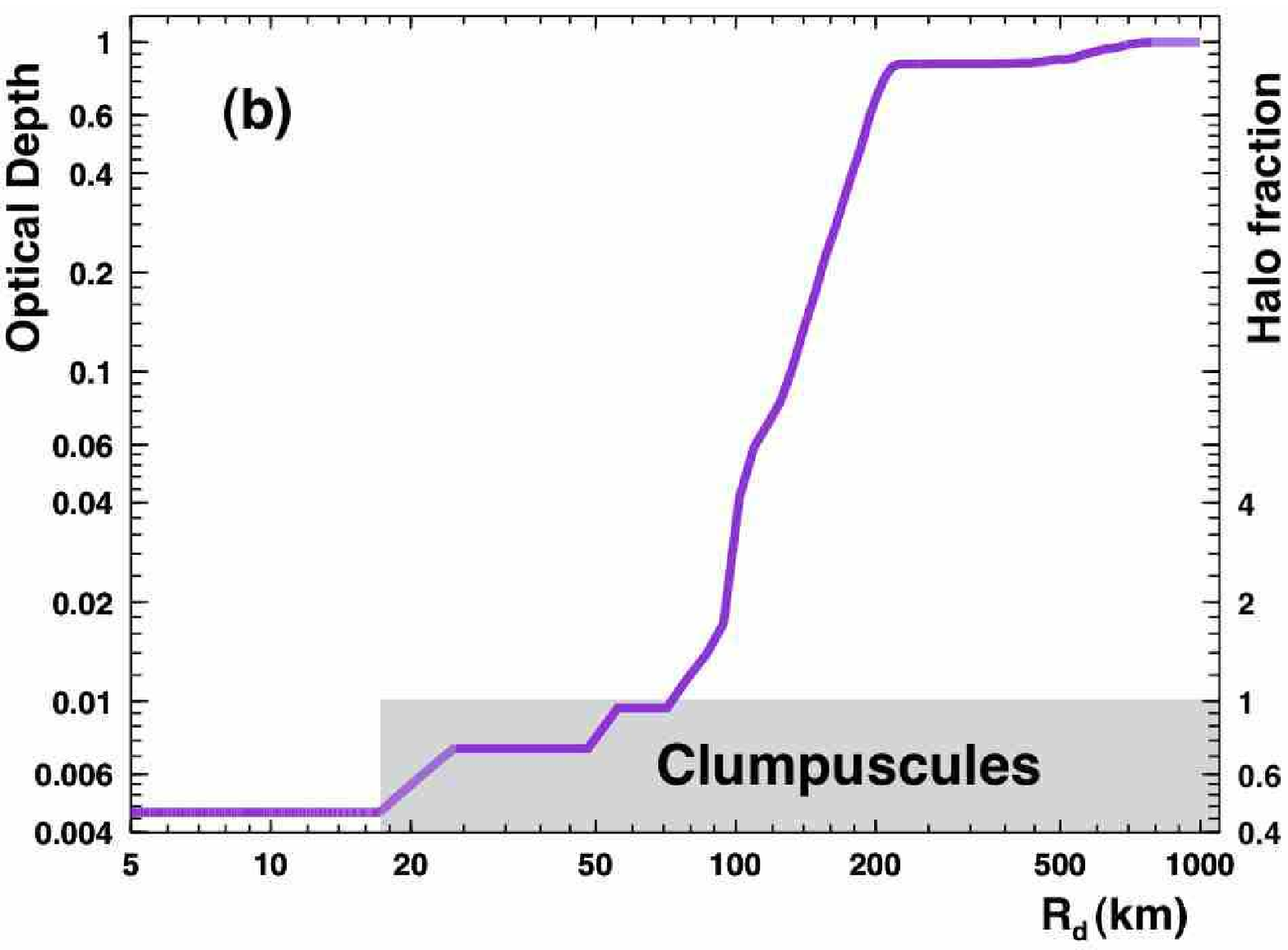}
}
\caption[]{\it
(a) $N_*(R_d)$, the number of SMC stars (or lines of sight) with no
turbulent structure of $R_{diff}(1.25\mu m)<R_d$
along the line of sight. The gray band shows the allowed $R_{diff}(1.25\mu m)$
region for clumpuscules (see text).
(b) The $95\%\,CL$ maximum optical depth of structures with $R_{diff}<R_d$
toward the SMC. The right scale gives the maximum contribution of
structures with $R_{diff}(1.25\mu m)<R_d$ to the Galactic halo (in fraction);
the gray zone gives the possible region for the hidden gas clumpuscules.
\label{Rdiff-SMC}}
\end{figure}

Here, the distance of the stars 
are all the same $z_0+z_1\sim z_1=62\, kpc$ (\cite{Szewczyk}).
Therefore the angular stellar radii
$\theta_s$ can be (roughly) estimated from the observed $J$ magnitude.
We used the EROS data (\cite{ErosLMCfinal},
\cite{Hamadache}, \cite{Rahal})
to distinguish between main sequence
and red giant branch stars
from the $R_{EROS}$ versus $(B_{EROS}-R_{EROS})$ color-magnitude diagram;
then the stellar type is obtained from the absolute magnitude
$M_J=J-5.\log(62000pc/10pc)$ and $\theta_s$ is derived from
Eq. (\ref{rvsk}) (see Fig. \ref{thetas}, left).
With these ingredients, we can estimate
from Eq. (\ref{rdiffmin})
$R_{diff}^{min}$ the lowest value of
$R_{diff}$ compatible with the measured modulation index $m$
for each stellar light-curve.
Figure \ref{Rdiff-SMC}(a) shows the cumulative distribution of
this variable $R_{diff}^{min}$,
that is $N_*(R_d)$, the number of stars whose line of sight do not cross
a gaseous structure with $R_{diff}<R_d$.
The bimodal shape of this distribution is caused by the
prominence of the red giants in the monitored population:
scintillation is expected to be less contrasted for
red giant stars because of their large radius;
therefore, structures
with $R_{diff}\gtrsim 250\, km$ cannot induce detectable
modulation in the red giant light-curves,
and the $R_{diff}^{min}$ values are lower than $250\, km$.

The distribution vanishes beyond $R_d\sim 800\, km$ because 
our limited resolution on the main sequence stars
(a few $\%$) prevents us from detecting any scintillation of
gaseous structures
with $R_{diff}>800\, km$.
The possible $R_{diff}$ domain for the hidden gas clumpuscules expected
from the model of (\cite{fractal1} 1994; \cite{fractal2}) and their maximum
contribution to the optical depth are indicated by the gray zone;
the minimum $R_{diff}$ ($\sim 17\, km$ at $\lambda =1.25\mu m$)
for these objects is estimated from Eq. (\ref{expression-rdiff}) assuming
the clumpuscule's outer scale is $30\, AU$ and considering
the maximum possible value of the density
fluctuation, mathematically limited by the maximum density
($\sigma_{3n}^{max}<n_{max}=10^{10} cm^{-3}$)(\cite{fractal1} 1994; \cite{fractal2}).

From the distribution of $N_*(R_d)$, one can infer limits on the
scintillation optical depth $\tau_{1.25\mu m}(R_d)$ as it is defined
in Sect. \ref{sec:optdepth}.
Indeed $N_*(R_d)$ is the number of lines of sights (l.o.s.) that do not cross
structures with $R_{diff}<R_d$.
Defining $N_{behind}$ as the total number of monitored
l.o.s. through the nebula (all monitored
stars in the present case because we are searching for invisible gas),
the upper limit on the optical depth $\tau_{1.25\mu m}(R_d)$
is the upper limit on the ratio
$p = \frac{N_{behind}-N_*(R_d)}{N_{behind}}$.
The $95\%$ statistical upper limit on $p$ is given by the classical confidence
interval {\it i.e.}
\footnote{the first formula gives the
upper limit of the $90\%\, CL$ interval for the $p$ value
(see classical texbooks like \cite{Ventsel}); therefore
the probability that the true $p$ value is higher than
the upper limit of this interval is $5\%$.}:
\begin{itemize}
\item
if $N_*(R_d)$ and $N_{behind}-N_*(R_d)$ are both larger than 4, then
\begin{equation}
\tau_{\lambda}(R_d) < p + 1.643 \sqrt{\frac{p(1-p)}{N_{behind}}};
\label{limtau}
\end{equation}
\item
if $N_*(R_d)\le 4$, then
\begin{equation}
\tau_{\lambda}(R_d) < \frac{N_{behind}-N_{95\%}(R_d)}{N_{behind}},
\end{equation}
where $N_{95\%}(R_d)$ is the $95\% C.L.$ Poissonian lower limit
on $N_*(R_d)$;
\item
if $N_{behind}-N_*(R_d)\le 4$, then
\begin{equation}
\tau_{\lambda}(R_d) < \frac{[N_{behind}-N_*(R_d)]_{95\%}}{N_{behind}},
\end{equation}
where $[N_{behind}-N_*(R_d)]_{95\%}$ is the $95\% C.L.$ Poissonian
upper limit on $N_{behind}-N_*(R_d)$.
\end{itemize}
As an example, the upper limit on the optical depth for structures
with $R_{diff}<400\, km$ is obtained from the value of $N_*(400\, km)\sim 100$;
as $N_{behind}=691$, it comes $p \sim 0.86$.
For this example, the $95\%\, CL$ upper limit on
$\tau_{1.25\mu m}(400\, km)$
is found to be $0.88$ from Eq. (\ref{limtau}).

The expected optical depth is proportional to the total mass of gas.
In \cite{fractal1} (1994) the clumpuscules are
expected to cover less than $\sim 1\%$ of the sky; this means that the maximum
optical depth should be $\sim 0.01$ assuming a Galactic halo
completely made of gaseous clumpuscules. Therefore,
we can interpret our optical depth limit as the upper limit of
the contribution of turbulent gaseous structures with $R_{diff}(1.25\mu m)<R_d$,
expressed in fraction of the halo
(right scale in Fig. \ref{Rdiff-SMC}(b)).

Our upper limit does not yet seriously constrain the model with clumpuscules,
but we can extrapolate these results to define a strategy
to reach a significant sensitivity.
Our monitored stellar population is dominated by red giant stars that
would give a lower scintillation signal than smaller stars from
the main sequence. This could only be compensated by achieving an
excellent photometric precision on the red giants.
An easier way to improve a hypothetic signal would be to
use $V$ passband instead of $J$.
In our test,
the use of $J$ filter was imposed by the choice of the SOFI
detector, the only one available with a fast readout.
According to the ESO exposure time calculator (\cite{exposure}),
the same precision we obtained in $J$ for the
red giants can be reached for $A0$ stars in $V$ with the same
exposure time (10s).
Therefore,
we can extrapolate that an exposure of $\sim 10^6 star\times hour$
(about 100 times more than our test) obtained with 
the same type of telescope (NTT) using the $V$
passband (around the maximum of stellar emission) should
provide enough sensitivity to significantly
constrain the turbulent gas component of the Galactic halo.

\section{Limits on the gas structuration in the nebulae}
For our study of known nebulae, the distance $z_0$ to the gas
is known, but not the star-by-star $z_1$ distances.
We first make the hypothesis that the stellar population behind
the clouds (whose light is absorbed and diffracted)
is the same as the population which is not -- or much less -- obscured
(the so-called control population, see Fig. \ref{mask}).
Toward B68
the red giant stars are distinguished from the main sequence stars
with our own $J$ image, also taken
with the NTT-SOFI detector, through the $K_s$ versus $(J-K_s)$ diagram.
Using Eq. (\ref{rvsk}), we established the $K_s$ to $\theta_s^{max}$
relation, providing the maximum angular radius of a
main sequence or red giant star located beyond $4\, kpc$ with $K_s$ apparent
magnitude (Fig. \ref{thetas} right).
We conservatively use this angular radius in Eq. (\ref{rdiffmin})
to estimate $R_{diff}^{min}$.
The study of the population behind the dust is complicated by the fact that
the stars are obscured.
In the case of B68 we were able to correct the
apparent magnitudes for the absorption in $K_s$ band;
we used the $A_V$ absorption map from (\cite{Alves}) and the
relation $A_K/A_V=0.089$ (\cite{Ojha}, \cite{Glass}) to extract $A_K$
and deduce the dereddened magnitudes $K_s$.
Using the relation of Fig. \ref{thetas} (right) with these corrected
magnitudes, we extract $\theta_s^{max}$ for each star behind the dust.
\begin{figure}[h]
\centering
\parbox{8cm}{
\includegraphics[width=8cm,height=6cm]{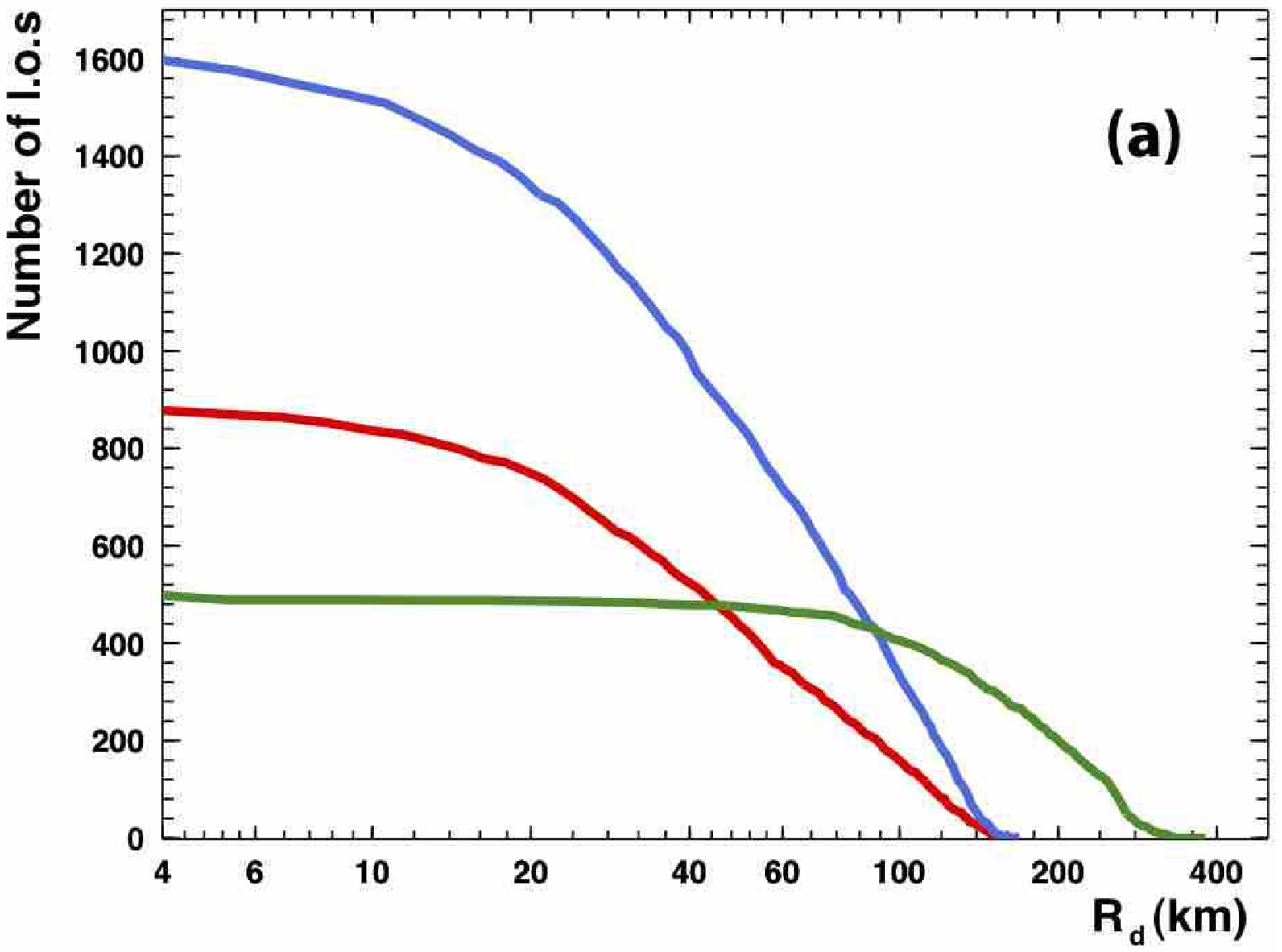}
\includegraphics[width=8cm,height=6cm]{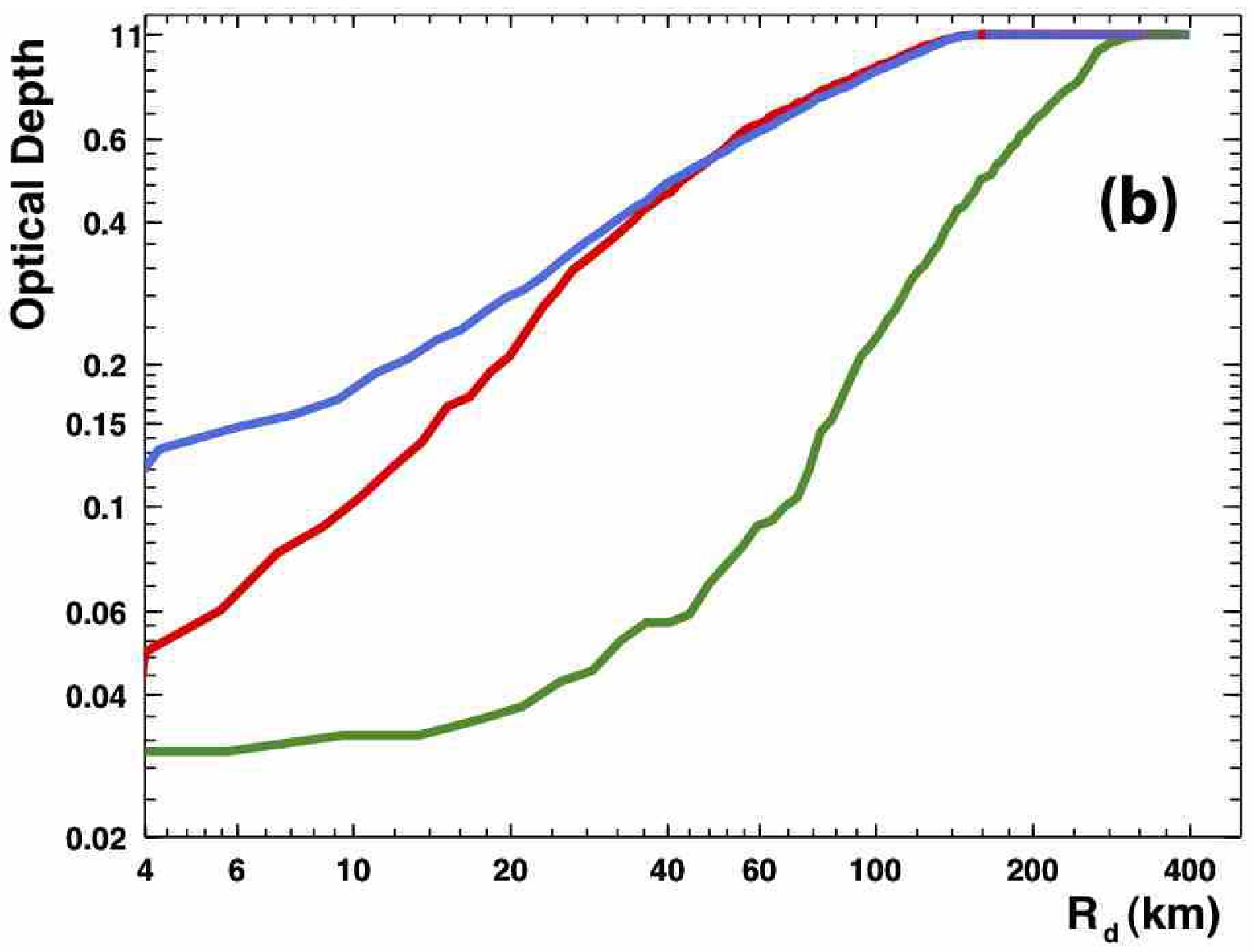}
}
\caption[]{\it
(a) $N_*(R_d)$, the number of directions with no turbulent structure
of $R_{diff}(2.16\mu m)<R_{d}$
along the line of sight toward the obscured regions of
B68 (green), cb131 (blue) and toward the Circinus nebula (red).\\
(b) The $95\%\,CL$ maximum optical depth of structures
with $R_{diff}(2.16\mu m)<R_{d}$.
\label{Rdiff-neb}}
\end{figure}

Because there is no absorption map for bc131 and the Circinus nebula,
we conservatively used the highest possible value of
$\theta_s^{max}\sim 13 \times \theta(R_{\odot}\ at\ 10 kpc)$
(corresponding to a $B3$ type star at $4.5\, kpc$).

Figure \ref{Rdiff-neb} shows the cumulative $R_{diff}^{min}$
distributions and the upper limits on $\tau_{2.16\mu m}(R_d)$
for the obscured regions of B68 and bc131
(the search regions from Fig. \ref{mask})
and over the complete field of the Circinus nebula,
which has indistinct boundaries.
The best limits are naturally obtained toward B68, as a consequence of
the better knowledge of the stellar sizes.
The shape of the cumulative $R_{diff}^{min}$ distributions differs
from the one obtained toward the SMC,
because the contribution of (big) red giant stars is smaller.

We can now interpret our limits on $R_{diff}$
to constrain the global structure
of the nebulae, and specifically put limits on
the existence and the structure of local turbulent dense cores
(cells) within the nebulae (see \cite{Lada} and \cite{Racca}).
\begin{itemize}
\item{\bf Probing the global structure?}\\
If the nebula is a simple object described by a single Kolmogorov
turbulence law, characterized by an outer scale of $L_z\sim 17000\, AU$
(maximum depth of gas crossed by the light in B68),
then the minimum expected value of $R_{diff}$ is deduced from
Eq. (\ref{expression-rdiff}), where $\sigma_{3n}$ is limited by $n_{max}$
the maximum molecular density.
This maximum is estimated to be $2.61\times 10^{5} cm^{-3}$ for B68
(\cite{Hotzel} 2002).
Using this constraint and the maximum outer scale value 
in (\ref{expression-rdiff}) gives for B68
\begin{eqnarray}
&&\!\!\! R_{diff}(2.16\mu m)> \nonumber \\
&&\!\!\! 263\, km
\left[\frac{2.16\mu m}{1\mu m}\right]^{\frac{6}{5}}
\left[\frac{17000\, AU}{10\, AU}\right]^{-\frac{1}{5}}\!
\left[\frac{2.61\!\times\! 10^{5}\, cm^{-3}}{10^{9}\, cm^{-3}}\right]^{-\frac{6}{5}} \nonumber \\
&& >3.\times 10^6 km.
\end{eqnarray}
For cb131, the same calculation based on data from (\cite{Bacmann} 2000)
gives $R_{diff}(2.16\mu m)>4.2\times 10^6 km$.
These large diffusion radii, which are owing to the weakness of the power spectrum at
small scales, are much too large to induce any observable
scintillation effect on any type of stars.
Indeed, because $\theta_s>\theta(R_{\odot}\ at\ 10 kpc)$, 
Eq. (\ref{contparam}) gives $R_s/R_{ref}>3100.$ at $\lambda=2.16\mu m$,
and the expected modulation index is completely negligible
(out of scale in Fig. \ref{modindex}).
\item{\bf Probing local sub-structures?}\\
In contrast, local turbulent dense cores with much smaller
diffusion radii could potentially produce observable scintillation. 
Therefore our upper limits on scintillation can be interpreted as upper
limits on the existence of turbulent cells with $R_{diff}(2.16\mu m)<350\, km$
within the volume of the nebula.
Using Eq. (\ref{expression-rdiff}), it can be interpreted as limits
on the distribution of the product
\begin{equation}
\left[\frac{\sigma_{3n}}{10^{9}\, cm^{-3}}\right]
\left[\frac{L_z}{10\ AU}\right]^{\frac{1}{6}}
= \left[\frac{R_{diff}(2.16\mu m)}{663\, km}\right]^{-\frac{5}{6}}
\end{equation}
along the lines of sight;
thanks to the small exponent of $L_z$, a rough hypothesis
for the core size would allow one to
extract upper limits on the frequency of
cores with a density dispersion higher than a given value $\sigma_{3n}$.

No conclusion can be infered for
structures with $R_{diff}>350\, km$, because they cannot produce a detectable
scintillation
in our sample because of our limited photometric precision on the small stars.
\end{itemize}
\section{Conclusions and perspectives}
The aim of the test was to study the feasibility of a systematic
search for scintillation. We were lucky enough to find a
stochastic variable light-curve that is compatible with a scintillation
effect; but considering the low probability of such an event, which is
related to the low density of the nebulae,
a program of synchroneous
multicolor and multi-epoch observations is necessary to obtain a
convincing signature of the effect.

From our search for invisible gas toward the SMC, we conclude that
an ambitious program using a wide field, fast readout camera at
the focal plane of a $>4m$ telescope should either discover turbulent gas
in the halo, or exclude this type of hidden baryonic matter.
With such a setup, significant results should be obtained with
an exposure of $\sim 10^6 star\times hour$ in $V$ passband.

The hardware and software techniques required for scintillation searches
are available just now, and a reasonably priced dedicated project could be
operational within a few years.
Alternatives under study are the use of the data from the
LSST project and from the GAIA mission.
If a scintillation indication is found in the future, one will have
to consider a much more ambitious project involving synchronized telescopes,
a few thousand kilometers apart.
Such a project would allow one to temporally
and spatially sample an interference pattern, unambiguously providing
the refractive length scale $R_{ref}$, the speed, and
the dynamics of the scattering medium.

\begin{acknowledgements}
We are grateful to the IPM for supporting F. Habibi during his stay
in Tehran.
We thank Prof. P. Schwemling for his help in the
determination of the variable star parameters, J-F. Glicenstein, F.
Cavalier and P. Hello for their participation to discussions.
We thank Prof. J-F. Alves for providing us with the B68 absorption map.
We are grateful to the referee for his constructive
remarks that allowed us to significantly improve this article.
\end{acknowledgements}

\appendix
\section{Connection between the diffusion radius and the gas structuration}

We consider a gaseous medium characterized by a cell with size
($L_x,L_y,L_z$). An electromagnetic plane wave propagating along $z$
and crossing the medium is distorted. The distorsion is caused by the variation
of the column density along the propagation of the light and can be
described by a two-dimensional phase delay $\phi(x,y)$. We
use the phase structure function to
characterize the variations of the phase as follows:
\begin{eqnarray}
\label{dphi}
D_{\phi}(x,y) & = & <(\phi(x'+x,y'+y)-\phi(x',y'))^2>      \\
& = & 2 [<\phi^2(x',y')> - <\phi(x'+x,y'+y) \phi(x',y')>]  \nonumber \\
& = & 2 [\xi_{\phi}(0,0) - \xi_{\phi}(x,y)],           \nonumber
\end{eqnarray}
where $\xi_{\phi}(x,y)$ is the correlation function of the screen
phase and is related to the phase spectral density
$S_{\phi}(q_x,q_y)$ (power spectrum per volume unit) through the
Fourier transform as
\begin{eqnarray}
\xi_{\phi}(x,y)  &=& \int\!\! \int  S_{\phi}(q_x,q_y) e^{2\pi i(x q_x +
y q_y)} dq_x dq_y. \label{invft}
\end{eqnarray}
Substituting equation (\ref{invft}) in equation (\ref{dphi}) we obtain
\begin{eqnarray}
D_{\phi}(x,y) & = & 2 \int\!\! \int S_{\phi}(q_x,q_y) (1-e^{2\pi i(x q_x
+ y q_y)}) dq_x dq_y. \label{dphispec}
\end{eqnarray}

The relation between the two-dimensional spectrum of the phase,
$S_{\phi}(q_x,q_y)$ and the three-dimensional spectrum of the number
density fluctuations has been derived for a plasma by Lovelace
(\cite{lovelace1}, \cite{lovelace2}).
Here we extend the concept to the optical wavelength for a medium
of molecular gas:
\begin{eqnarray}
S_{\phi}(q_x,q_y) &=& 2\pi L_z \left(\frac{(2\pi)^2 \alpha}{\lambda}\right)^2
S_{3\delta n}(q_x,q_y,q_z=0), \label{lovelace}
\end{eqnarray}
where $S_{3\delta n}(q_x,q_y,q_z)$ is the spectrum of the density
fluctuations of the molecular gas in three dimensions
($\delta n= n-<n>$), $\alpha$ is the average polarizability of the molecules and
$L_z$ is the thickness of the medium along the line of sight.

Assuming the gaseous medium to be isotropically turbulent,
the 3D spectral density obeys a power
law relation within the turbulence inertial range:
\begin{eqnarray}
S_{3\delta n}(q_x,q_y,q_z) &=& C_n^2 q^{-\beta} \,\,\,\,\,\,\,\,\,
L_{out}^{-1} < q < L_{in}^{-1}\, ,
\label{scalings}
\end{eqnarray}
where $q=\sqrt{q_x^2+q_y^2+q_z^2}\ $, $\beta = 11/3$ is taken for the
Kolmogorov turbulence, $C_n^2$ is the
turbulence strength parameter and $L_{out}$ and $L_{in}$ are 
the outer and inner scales of the turbulence respectively. By substituting
equations (\ref{scalings}) and (\ref{lovelace}) in equation
(\ref{dphispec}), we compute the phase structure function in the polar
coordinate system:
\begin{eqnarray}
D_{\phi}(r) &=& 2 C_n^2 (2\pi) L_z (\frac{(2\pi)^2
\alpha}{\lambda})^2 \int_0^{\infty}\!\! \int_0^{2\pi}\! q^{-\beta}
(1-e^{2\pi i r q cos \theta}) d\theta q dq. \nonumber
\end{eqnarray}
The integration results in
\begin{eqnarray}
D_{\phi}(r) &=& 2 C_n^2 (2\pi)^{\beta} f(\beta) L_z \left(\frac{(2\pi)^2
\alpha}{\lambda}\right)^2 r^{\beta-2}. \label{dphir}
\end{eqnarray}
where
\begin{eqnarray}
f(\beta) &=& \int_0^{\infty} s^{1-\beta} (1-J_0(s)) ds \,\,=\,\, \frac{2^{-\beta}\, \beta \,
  \Gamma(-\beta/2)}{\Gamma(\beta/2)}. \nonumber
\end{eqnarray}
For the Kolmogorov turbulence $f(\beta = 11/3) \sim$ 1.118.

We define the diffusion radius $R_{diff}$ as the transverse scale
for which $D_{\phi}(R_{diff}) = 1~rad$.
$R_{diff}$ is directly obtained from equation (\ref{dphir}):
\begin{eqnarray}
R_{diff} &=& [2 C_n^2 (2\pi)^{\beta+4} f(\beta) L_z \alpha^2
\lambda^{-2}]^{1/(2-\beta)}. \label{rdiff}
\end{eqnarray}
We will now link the turbulence parameter $C_n^2$ to the
dispersion of the density fluctuation.
From Parseval's theorem the dispersion of the density fluctuation
is equal to the auto-correlation function at origin:
\begin{equation}
\sigma_{3n}^2 = \frac{1}{L_xL_yL_z}\int \delta_n^2 dx dy dz = \xi(0).
\end{equation}
On the other hand the auto-correlation is equal to the integration
over the spectrum in the Fourier space
\begin{equation}
\xi(0) = \int S_{3\delta n}(q) d^3q.
\end{equation}
Using equation (\ref{scalings}), the dispersion of the density
fluctuations is related to $C_n^2$ as
\begin{eqnarray}
\sigma_{3n}^2 &=& \int_{L_{out}^{-1}}^{L_{in}^{-1}} C_n^2 q^{-\beta}
(4\pi) q^2 dq \,\,=\,\, \frac{4\pi C_n^2}{\beta-3}
(L_{out}^{\beta-3} - L_{in}^{\beta-3}).
\end{eqnarray}
Assuming $L_z = L_{out} \gg L_{in}$, for $\beta$ = 11/3, we estimate
$C_n^2$ as
\begin{eqnarray}
C_n^2 &=& \frac{\sigma_{3n}^2}{6 \pi L_z^{2/3}}\,\,\,. \label{cn2}
\end{eqnarray}
Finally by substituting equation (\ref{cn2}) in (\ref{rdiff}),
$R_{diff}$ is obtained in terms of the cloud's parameters and the
wavelength as
\begin{eqnarray}
R_{diff}\!=\! 231.1\,km
\left[\frac{\alpha}{\alpha_{H_2}}\right]^{-\frac{6}{5}} \!
\left[\frac{\lambda}{1 \mu m}\right]^{\frac{6}{5}} \!
\left[\frac{L_z}{10 AU}\right]^{-\frac{1}{5}} \!
\left[\frac{\sigma_{3n}}{10^9 cm^{-3}}\right]^{-\frac{6}{5}}\!\!\!\!\!\! ,
\end{eqnarray}
where $\alpha_{H_2}=0.802\times 10^{-24} cm^3$ is the polarisability
of molecule ${\rm H_2}$.

\section{Limitations of the photometric precision}
In our photometric optimization, we found that
the photometric precision on unblended stars
was limited by several factors.
First, the PSF of very bright stars significantly differs from
the ideal Gaussian
and the photometric precision on these stars is
seriously affected.
Second, the best photometric precision does not follow the naive
expectation assuming only Poissonian noise.
For stars fainter than $K_s=14.6$ (or $J=16.6$)
we were able to reproduce the behavior of the PSF fit $\chi^2$
and of the fitted flux uncertainties
by assuming that the relative uncertainty $\sigma_{i,j}$ on a pixel content
results from the combination of the Poissonian
fluctuations of the number of photoelectrons $N_{i,j}^{\gamma e}$,
and of a systematic uncertainty ({\it i.e.} not changing with time, but depending
on the pixel) due to the flat-fielding procedure:
\begin{equation}
\sigma_{i,j}^2=\frac{1}{N_{i,j}^{\gamma e}}+\left[\frac{\Delta C_{i,j}}{C_{i,j}}\right]^2,
\end{equation}
where $N_{i,j}^{\gamma e}$ is the {\it total} number of photoelectrons (from
the fitted star and the -- usually dominant -- sky background), and
$\frac{\Delta C_{i,j}}{C_{i,j}}$
is the uncertainty of the flat-field coefficient for pixel $(i,j)$.
From the comparison of two flat-fields taken at different epochs, we
estimated that $\frac{\Delta C_{i,j}}{C_{i,j}}\sim 0.63\%$ (both in $K_S$
and in $J$).
A third source of noise in $J$ comes from a residual fringing;
we measured that fringes contribute to a $\sim 40\%$
increase of the Poissonian fluctuation $\sqrt{N_{i,j}^{\gamma e}}$
in a domain size comparable to
the PSF width because of their small spatial scale.

We conclude that the contribution of the Poissonian
noise dominates the error when the flux from the star plus the total sky
background within the PSF fitting domain is
small ($\sim 2000\ ADU=10600 \gamma e$). 
For brighter stars or for stars located in a dusty environment
(producing a large sky IR background), the second -- systematic -- term
dominates.
\end{document}